\documentstyle[epsfig]{mn}
\begin{document}

\title
[Ultraluminous infrared galaxies at high redshift]
{Ultraluminous infrared galaxies at high redshift: 
their position on the Madau plot and their fate} 
\author[Neil Trentham, A.\,W. Blain \& Jeff Goldader]
{
Neil Trentham,$^1$ A.\,W. Blain$^2$ and Jeff Goldader$^3$\\
$^1$ Institute of Astronomy, Madingley Road, Cambridge, CB3 0EZ.\\
$^2$ Cavendish Laboratory, Madingley Road, Cambridge, CB3 0HE.\\
$^3$ University of Pennsylvania, Department of Physics and Astronomy, 
209 S.~33rd St., Philadelphia, PA 19104, USA.
}
\maketitle 

\begin{abstract} 
A major recent development in extragalactic astronomy has been the discovery
of a population of galaxies that are luminous at submillimetre wavelengths. 
Estimates of their spectral energy distributions suggest that these 
galaxies are the high-redshift analogues of the ultraluminous infrared galaxies 
(ULIRGs) observed locally. Here we investigate the implications for the galaxy 
formation problem if this is indeed the case. We identify plausible (but 
non-unique) redshift-dependent galaxy luminosity functions that are consistent 
with both source counts at 2800, 850, 450 and 175\,$\mu$m, and far-infrared 
background radiation intensities at 850, 240 and 140\,$\mu$m. In all our models, 
most of the submillimetre-luminous sources are distant galaxies with high 
bolometric luminosities $\geq 10^{12}$\,L$_{\odot}$. As for many local ULIRGs, 
it is not possible to determine whether these luminous galaxies 
are powered by starbursts, like the local galaxy Arp 220, or by 
active galactic nuclei (AGN), like the local galaxy Markarian 231. We investigate 
both possibilities. If the submillimetre-luminous galaxies are all starbursts, 
then we predict that the fraction of the cosmic star-formation rate in these 
objects is large, but does not necessarily dominate the star-formation rate 
of the Universe -- the Madau plot -- at any redshift. Only a few per cent by 
mass of the present-epoch spheroidal stellar population would have been
formed in such a population of star-forming galaxies, consistent with the 
constraints on the number of galaxies with old stellar populations in the field at 
low and intermediate redshifts derived from $K$-band surveys. 
If the 
submillimetre-luminous galaxies are all powered by AGN, then the comoving 
density of supermassive black holes onto which material is accreting at high 
redshift probably equals no more than a few
per cent of the local density of massive 
dark objects. 
\end{abstract} 


\section{Introduction} 

Recent observations with the Submillimetre Common-User Bolometer Array 
(SCUBA; Holland et al.\ 1999) on the James Clerk Maxwell Telescope have 
highlighted the presence of a number of submillimetre-luminous galaxies 
(Smail, Ivison \& Blain 1997; Holland et al.\ 1998; Barger et al.\ 1998a; 
Hughes et al.\ 1998; Eales et al.\ 1998). To date about forty 
sources have been found. These measurements confirm earlier suggestions 
(Blain \& Longair 1993; Eales \& Edmunds 1996, 1997) that submillimetre-wave 
observations will provide an important probe of cosmology.

The optical counterparts of these SCUBA sources are faint, most with $I > 20$ 
(Smail et al. 1998). They are presumably not low-redshift sources: the best 
studied is SMM\,J02399$-$0136 (Ivison et al.\ 1998), a dust-enshrouded AGN at 
$z=2.8$, perhaps similar to IRAS F10214+4724 (Rowan-Robinson et al.\ 1991; 
Lacy et al.\ 1998).  The recent detection of molecular gas in SMM\,J02399$-$0136 
(Frayer et al.\ 1998) also suggests that these galaxies are similar.
 
What is the nature of these SCUBA-selected sources? In particular are they 
similar to the ultraluminous infrared galaxies (ULIRGs) observed locally (e.g. 
Sanders \& Mirabel 1996)? They have the same submillimetre-wave flux densities 
as would have local ULIRGs seen at high redshift (Barger et al.\ 1998a). They also have 
similar optical colours (Smail et al.\ 1998) to two of the three local ULIRGs
studied at ultraviolet wavelengths by Trentham, Kormendy \& Sanders
(1998), were those galaxies to be seen at high redshift. If this interpretation is 
correct, then we might hope to be able to use the wealth of observational 
information available for local ULIRGs to help to understand the properties of the 
SCUBA sources and their relevance to cosmology.  

Whether the SCUBA sources are dust-enshrouded AGNs, like Markarian 231, or 
dust-enshrouded starbursts, like Arp 220, is an important question. 
This is difficult to determine for even local ULIRGs, because there are up to 
several hundred magnitudes of extinction along our lines of sight to the galaxy 
cores at optical wavelengths. For local ULIRGs, simple conclusions can be 
drawn based on the general form of the spectral energy distributions (SEDs) 
(e.g.~Sanders et al.\ 1988a); however, this information is not available for the 
SCUBA sources. Recently, new more detailed methods based on mid-infrared 
spectroscopic diagnostics (Lutz et al.\ 1996, Genzel et al.\ 1998) have been used 
to resolve this question for the most local ULIRGs. 
 
We construct redshift-dependent luminosity functions that are consistent with
all the counts and backgrounds from a number of recent surveys at far-infrared 
and millimetre/submillimetre wavelengths. We use non-evolving 
Gaussian luminosity functions over specified redshift ranges.  These are the 
simplest luminosity functions that we could adopt, without the results 
being dependent on the properties of low-luminosity galaxies that are not 
probed by the SCUBA observations. Contributions from these low-luminosity 
galaxies are important when computing backgrounds, although they contribute 
insignificantly to the counts. The parameterization used to compare models 
with observations reflects this fact. Imposing all the 
constraints simultaneously does limit the possible form of the luminosity 
function and we identify three plausible models.

Given the redshift-dependent luminosity functions of these models, we 
investigate the properties to the individual sources using observations of their 
local ultraluminous counterparts, and derive the cosmological implications of 
the SCUBA galaxies under both the starburst and AGN interpretation.

We first investigate the possibility that the SCUBA sources are high-redshift 
star-forming galaxies. The starburst models of Leitherer \& 
Heckman (1995) are used to convert far-infrared luminosities to star-formation 
rates; a transformation that is uncertain by more than an order of magnitude.
We then place the SCUBA galaxies on the Madau plot, which relates the 
cosmic star formation rate to the cosmic epoch (Madau et al.\ 1996; Madau, 
Della Valle \& Panagia 1998), 
and predict the total density of local stars produced in such 
objects. Most of the star formation in the SCUBA sources is observed through
huge amounts of internal extinction, and so will not be included in global
star-formation rates that are computed using the optically selected samples that 
are normally used to construct the Madau plot.  Indeed Hughes et al.\ (1998) 
show that deriving a star-formation rate from a rest-frame ultraviolet flux 
results in a value that is more than an order of magnitude too low.
The SCUBA sources are thus not accounted for in the existing samples that are 
used to construct the Madau plot.

We then investigate the possibility that the SCUBA sources are high-redshift
dust-enshrouded AGN, as discussed by Haehnelt, Natarajan \& Rees (1998) 
and Almaini, Lawrence \& Boyle (1999). The three distant 
submillimetre-luminous galaxies that have been studied in detail -- 
APM\,08279+5255 (Irwin et al. 1998; Lewis et al. 1998; Downes et al. submitted), 
SMM\,J02399$-$0136 and IRAS F10214+4724 -- all contain powerful AGN. If all 
the sources derive their bolometric luminosity from AGN that heat their dust 
shrouds radiatively by accretion onto a massive black hole, then we can use our 
redshift-dependent luminosity functions to
compute the comoving integrated mass density of these black holes.

We find plausible (although not unique) scenarios that are consistent with all 
the observations and present their cosmological implications. Finally, we 
highlight future work which may help to distinguish between the various 
scenarios. Throughout this paper we assume an Einstein--de Sitter world model 
with $H_0 = 50$\,km\,s$^{-1}$\,Mpc$^{-1}$.

\section{Far-infrared/submillimetre-wave counts and backgrounds
and the ULIRG luminosity function}

There have been many recent measurements of the far-infrared and
submillimetre source counts and backgrounds at a number of wavelengths:
see Table 1.  This has been a recent field of substantial activity 
because of new instrumentation, both SCUBA and {\it ISO}. If we assume that
the SED at far-infrared and submillimeter
wavelengths is known for all the sources,
we can then use all these measurements in conjunction to constrain the
bivariate far-infrared luminosity--redshift function of these sources.

\begin{table*}
\caption{Far-infrared and submillimetre-wave surveys.
Lagache et al.~(1998) and Kawara et al.~(1997, 1998) obtain a different
calibration for the ISO 175-$\mu$m counts.  A discussion of calibration is
given in Lagache et al.~(1998). The detection limit in the 850-$\mu$m 
survey by Hughes et al. (1998) is about 2\,mJy: the count at 1\,mJy is 
derived from a source confusion analysis. See Blain et al.~(1999b) and 
Smail et al. (1999) for a 
direct sub-mJy 850-$\mu$m count. For complementary 
far-infrared/submillimetre-wave background measurements the reader 
is referred to Puget et al.~(1996), Guiderdoni et al.~(1997), Dwek et al.~(1998) 
and Fixsen et al.~(1998).}
{\vskip 0.75mm}
{$$\vbox{
\halign {\hfil #\hfil && \quad \hfil #\hfil \cr
\noalign{\hrule \medskip} 
Wavelength  & Counts /     & Flux Limit /& Background /         & 
Telescope/Instrument & Reference &\cr
            & deg$^{-2}$  & mJy        & nW m$^{-2}$ sr$^{-1}$  &
       &           &\cr 
\noalign{\smallskip \hrule \smallskip} 
\cr 
2.8\,mm & $<162$ & 3.55 & $-$ &  BIMA          & Wilner \& Wright (1997) &\cr 
\noalign{\smallskip}
850\,$\mu$m &  $2500\pm1400$ & 4  & $-$ &  JCMT/SCUBA & 
Smail et al.~(1997)&\cr
	    &  $1100\pm600$ & 8 & $-$ & & Holland et al.~(1998)&\cr 
	    &  $800^{+1100}_{-500}$ & 3 & $-$ & & Barger et al.~(1998a) & \cr
            &  $7000\pm3000$ & 1 & $-$ & & Hughes et al.~(1998) & \cr 
	    &  $1800\pm600$ & 2.8 & $-$ & & Eales et al.~(1998) &\cr 
\noalign{\smallskip}
850\,$\mu$m & $-$ & $-$ & $0.55 \pm 0.15$ & {\it COBE}/FIRAS & 
Fixsen et al.~(1998)
&\cr
\noalign{\smallskip}
450\,$\mu$m &  $< 1000$ & 80 & $-$ & JCMT/SCUBA & Smail et al.~(1997) &\cr
	    &  $<360$ & 75 & $-$ & & Barger et al.~(1998a) &\cr 
\noalign{\smallskip}
240\,$\mu$m & $-$ & $-$ & $17 \pm 4$ &  {\it COBE}/DIRBE + {\it IRAS}/ISSA & 
Schlegel et al.~(1998) &\cr 
            & $-$ & $-$ & $14 \pm 4$ & & Hauser et al.~(1998) &\cr 
\noalign{\smallskip}
175 $\mu$m &  $41 \pm 6$ & 150 & $-$ & {\it ISO}/ISOPHOT & Kawara et
al.~(1998) 
&\cr
	    & $98 \pm 15$ & 100 & $-$ & 	    & 
Lagache et al.~(1998) &\cr
\noalign{\smallskip}
140 $\mu$m & $-$ & $-$ & $32 \pm 13$ & {\it COBE}/DIRBE + {\it IRAS}/ISSA & 
Schlegel et al.~(1998) &\cr
	   & $-$ & $-$ & $24 \pm 12$ & & Hauser et al. (1998) &\cr
\noalign{\smallskip \hrule} 
\noalign{\smallskip}\cr}}$$}
\end{table*} 

We assume that the galaxies have the thermal SEDs of 
warm dust that is heated by an enshrouded starburst or AGN, and adopt a 
simple analytic form for the luminosity function between a minimum and 
maximum redshift. All the observational constraints on the model are imposed 
through integral functions, and so do not require unique solutions.  
Nevertheless, it is interesting to see which general classes of 
luminosity--redshift distributions are ruled out and why.

To ensure consistent definitions are used when matching models to
observations, first we present all the details of our computations.  Much of this 
will be familiar to many readers, but different authors define parameters in slightly 
different ways. Secondly, we point out some generic features of the comparison 
between our models and observations, in particular the results of requiring the 
models to be consistent with source counts at one wavelength and with the 
infrared background at another simultaneously. Finally, we isolate some 
plausible luminosity functions that are consistent with all the observations for 
further investigation.

We define a bivariate 60-$\mu$m luminosity-redshift function 
$\phi_z (L_{\rm 60})$, with units Mpc$^{-3}$\,${\rm L}_{\odot}^{-1}$, such that 
$\phi_z (L_{\rm 60}) \, {\rm d} L_{\rm 60}\, {\rm d}z$ is the total number density of 
galaxies with 60-$\mu$m luminosity between $L_{\rm 60}$ and 
$L_{\rm 60} + {\rm d} L_{\rm 60}$ with redshifts between $z$ and $z + {\rm d}z$. 
Some familiar analytic examples of this function are a 
Gaussian in $\log_{10} L_{\rm 60}$ with no luminosity or density evolution, 
\begin{equation} 
\phi_z (L_{\rm 60}) = C (1+z)^3 \exp \left[ - {1\over{2 \sigma^2}}  
\log_{10}^2 \left( {{L_{\rm 60}} \over {L_{\rm 60}^*}}\right) \right]   
{1\over{L_{\rm 60} \, \rm{ln} 10}},
\end{equation} 
or a Saunders et al.\ (1990) function, which is a power-law in $L_{\rm 60}$
with index $\alpha$ at the faint end and a Gaussian in $\log_{10} L_{\rm 60}$ at 
the bright end, and allows for density and luminosity
evolution,
\begin{eqnarray} 
\lefteqn{\nonumber 
\phi_z (L_{\rm 60}) = C(z)\, (1+z)^{3} 
\left( {{L_{\rm 60}} \over {L_{\rm 60}^* (z)}} \right)^{1 - \alpha} \times } \\ 
& & \>\>\>\>\>\>\>\>\>\>\>\>\>
 \exp \left[ - {1\over{2 \sigma^2}}
\log_{10}^2 \left( 1 + {{L_{\rm 60}} \over {L_{\rm 60}^* (z)}}\right) \right]
{1\over{L_{\rm 60} \, \rm{ln} 10}}.
\end{eqnarray} 
Implicit in both equations (1) and (2) above  
are a normalization constant $C$ (units Mpc$^{-3}$, which depends on
$z$ if there is density evolution), a characteristic
luminosity $L_{\rm 60}^{*}$ (which depends on $z$ if there is
luminosity evolution), and a Gaussian width $\sigma$.

Once we have specified $\phi_z (L_{\rm 60})$ for some
population of galaxies, we can compute their contribution to the 
cosmic infrared background at some frequency $\nu$:
\begin{equation} 
\nu {\rm I}_{\nu} = {{1}\over{4 \pi}} \int_{0}^{\infty} \int_{0}^{\infty}
{ {l_{\nu} (z,L_{\rm 60})}\over{4 \pi d_{\rm L} (z)^2}} \,\, \phi_z (L_{\rm 60})
\, \, {\rm d} L_{\rm 60}  \, \, { {{\rm d} V}\over{{\rm d} z}} \, \, {\rm d} z 
,
\end{equation}
in units of W\,m$^{-2}$\,sr$^{-1}$, where, 
\begin{equation} 
l_{\nu} (z,L_{\rm 60}) = \left[ { {\nu (1+z)} \over {\nu_{60}}}\right]^4
\!\! L_{\rm 60} 
{ {k_{\nu (1+z)}}\over{k_{\nu_{60}}}} \,
{ {\exp \left({{h \nu_{60}}\over{kT}} \right) - 1 }\over
{\exp \left[{{h \nu(1+z)}\over{k T}} \right] - 1 } },  
\end{equation}
and $\nu_{60}$ 
is the frequency corresponding to a wavelength of 60\,$\mu$m. 
$k_{{\nu}}$ is the
emissivity function of dust.  We follow Hughes (1996) and 
Blain et al.\ (1999a) in 
assuming a power law $k_{\nu} \sim {\nu^{1.5}}$, and so the dust emission 
spectrum at long wavelengths is a Raleigh--Jeans power-law with spectral 
index 3.5. We can also compute the number density of sources with flux density 
above some threshold $S_{\rm lim}$, which is measured in units of
W\,m$^{-2}$\,Hz$^{-1}$ (or Jy), 
\begin{equation} 
n(S_{\rm lim}) 
= {{1}\over{4 \pi}} \int_{0}^{\infty} \int_{L_{\rm lim} 
[S_{\rm lim}]}^{\infty} \>  
\phi_z (L_{\rm 60})
\, \, {\rm d} L_{\rm 60}  \, \, { {{\rm d} V}\over{{\rm d} z}} \, \, {\rm d} z
,
\end{equation}
in units of sr$^{-1}$ (or deg$^{-2}$), where, 
\begin{eqnarray} 
\lefteqn{\nonumber 
L_{\rm lim} [S_{\rm lim}] = 
4 \pi d_{\rm L} (z)^2 \, \, S_{\rm lim} \, \nu_{60} 
\left[ { {\nu_{60}} \over {\nu (1+z)} } \right]^{3} \times } \\
& & \>\>\>\>\>\>\>\>\>\>\>\>\>\>\>\>\>\>\>\>\>\>\>\>\>\>\>\>\>\>
\>\>\>\>\>\>
\, { {k_{\nu_{60}}} \over {k_{\nu(1+z)}} } \,
{ {\exp \left[{{h \nu (1+z)}\over{k T}} \right] - 1} \over
 {\exp \left({{h \nu_{60}}\over{kT}} \right) - 1 }}
.
\end{eqnarray} 
The quantities,
\begin{equation} 
d_{\rm L} (z) = 
{ {2 c}\over{H_0}} \left( 1 - {{1}\over{\sqrt{1+z}}} \right) (1+z), 
\end{equation} 
and,
\begin{equation} 
{ {{\rm d} V}\over{{\rm d} z}} =
16 \pi \, \left({ { c}\over{H_0}} \right)^3
\, (1+z)^{-{{3}\over {2}}} \, 
\left( 1 - {{1}\over{\sqrt{1+z}}} \right)^2,
\end{equation} 
are the luminosity distance and the
redshift-derivative of the volume element respectively (assuming
$\Omega_0 = 1$).

A few features of the comparison between observation and theory
are immediately apparent. Source counts constrain the integral of 
$\phi_z(L_{60})$ over redshift and luminosity above some redshift-dependent 
limit. However, measurements of the background light
constrain the integral of the product of luminosity $L_{60}$ and
$\phi_z(L_{60})$ over redshift and all luminosities. For a Saunders luminosity 
function with pure density evolution, we compare in Fig.\,1 the ratio
of the predicted 850-$\mu$m source count and 240-$\mu$m background 
as compared with the observed values listed in Table\,1 as a function of the 
characteristic luminosity $L_{*} \equiv L_{60}^{*}$. 
The minimum is produced by the different constraints on the integral of 
$\phi_z(L_{60})$ imposed by the source counts and the background 
measurement. 

Very low luminosity galaxies make a negligible contribution to the source counts, 
but a substantial one to the infrared background if $L_*$ is low in the 
Saunders function. Hence, a low normalization is required to explain the 
background data, but a high one is required to explain the source count data,
and so the ratio plotted in Fig.\,1 is very high at low $L_*$. Increasing $L_*$
increases the fraction of high-luminosity galaxies, and so lowers the ratio of the
normalizations required in the figure.  At very high $L_*$, the source counts are 
produced by a very small number of extremely luminous sources, which cause 
the background to be extremely high based on their luminosities alone. Hence, 
the ratio of the normalization begins to increase again, producing the minimum
in Fig.\,1.  If the flux threshold (lower limit) in the luminosity integral in
equation (5) is reduced, then the turn-up of the ratio at high luminosities is less 
marked, and disappears as the lower limit tends to zero.

It is intriguing that this minimum in the ratio occurs at a very high characteristic 
luminosity  $L_{*} = 10^{12}$\,L$_\odot$. Locally $L_{*} \sim 10^{9}$\,L$_\odot$, 
and so if the luminosity function of the SCUBA sources has a Saunders form, 
huge amounts of luminosity evolution are required to match the data. This was 
essentially one of the main conclusions of Blain et al.\ (1999a).

\begin{figure} 
\begin{center}
\vskip-2mm
\epsfig{file=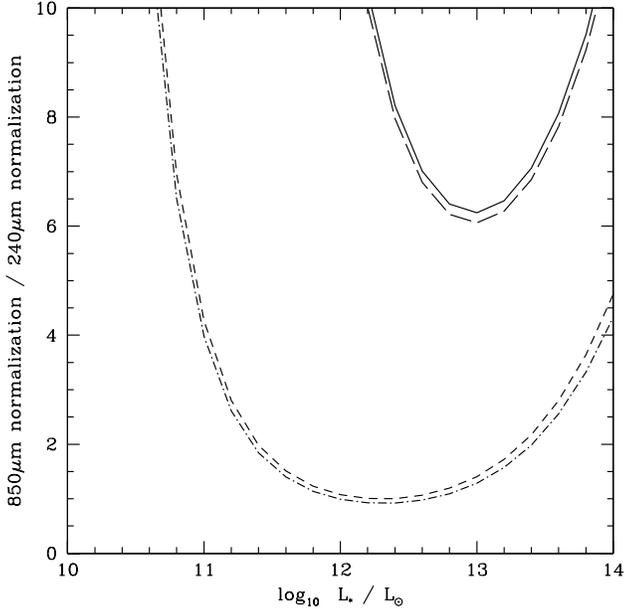, width=8.65cm}
\end{center}
\vskip-4mm
\caption{ 
The normalization obtained by fitting a Saunders function 
(equation 2) with density evolution parameterized by $\gamma$ to the 
COBE 240-$\mu$m background (Schlegel et al.\ 1998), relative to the 
normalization obtained by fitting a similar function to the SCUBA 
850-$\mu$m counts of Smail et al.\ (1997). A value of $\sigma = 0.724$, 
derived by Saunders et al.\ (1990), is assumed.  A dust emissivity index 
of 1.5 is assumed when converting luminosities measured at different 
wavelengths to restframe 60-$\mu$m luminosities. The four line styles 
represent the following parameter values: solid ($T=70$\,K, $\gamma = 0$),
short-dashed ($T=40$\,K, $\gamma = 0$), long-dashed ($T=70$\,K, 
$\gamma = 6$), and dot-dashed ($T=40$\,K, $\gamma = 6$). Locally,  
$L^{*}_{60} = 1.1 \times 10^9$\,L$_{\odot}$.
}
\end{figure} 

\begin{figure*}
\begin{minipage}{170mm}
{\vskip-3.5cm} 
\begin{center}
\epsfig{file=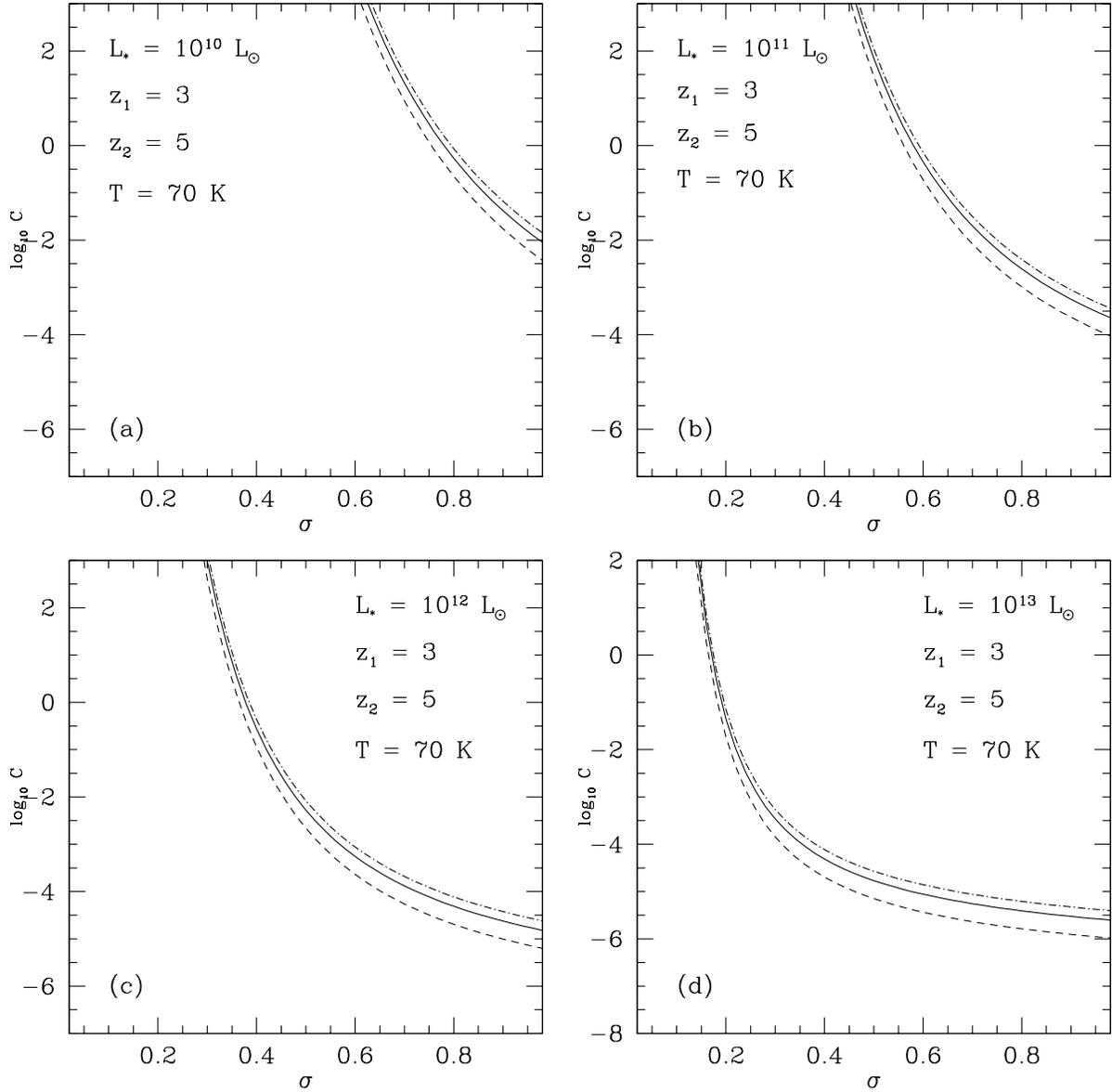, width=18.65cm}
\end{center}
{\vskip-7.2cm} 
\caption{
Combinations of the normalization constant $C$ and the width $\sigma$ for 
Gaussian luminosity functions with no density evolution that are consistent 
with the 850-$\mu$m counts.  The units of $C$ are Mpc$^{-3}$. All galaxies 
are assumed to have a temperature $T = 70$\,K. The comoving galaxy 
density $C$ is assumed to be constant between redshifts $z_1 = 3$ and 
$z_2 = 5$ and zero outside this range, (that is the upper and lower limits to 
the redshift integral in equation (5) are $z_1$ and $z_2$). The solid lines show 
fits to the mean value quoted by Smail et al.\ (1997). 
The  dotted-dashed lines are fitted to the +1$\sigma$ values. The dashed 
lines are fitted to the $-\sigma$ values: see Table 1. The four panels are 
plotted for different values of $L_{*}$.
}
\end{minipage}
\end{figure*}

\begin{figure*}
\begin{minipage}{170mm}
{\vskip-3.5cm}
\begin{center}
\epsfig{file=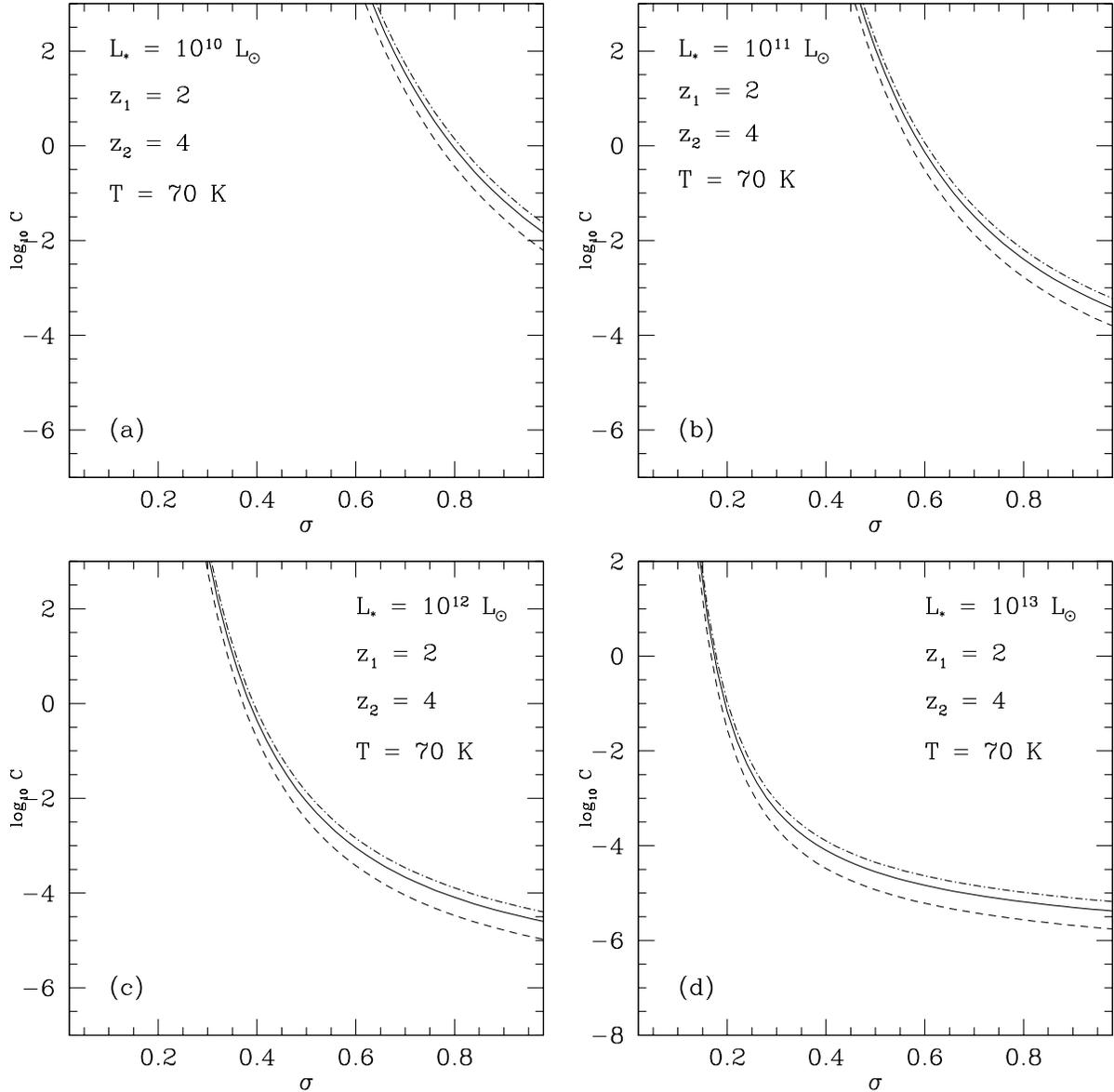, width=18.65cm}
\end{center}
{\vskip-7.2cm}
\caption{As Figure 2, but for $z_1=2$ and $z_2=4$.}
\end{minipage}
\end{figure*}

\begin{figure*}
\begin{minipage}{170mm}
{\vskip-3.5cm}
\begin{center}
\epsfig{file=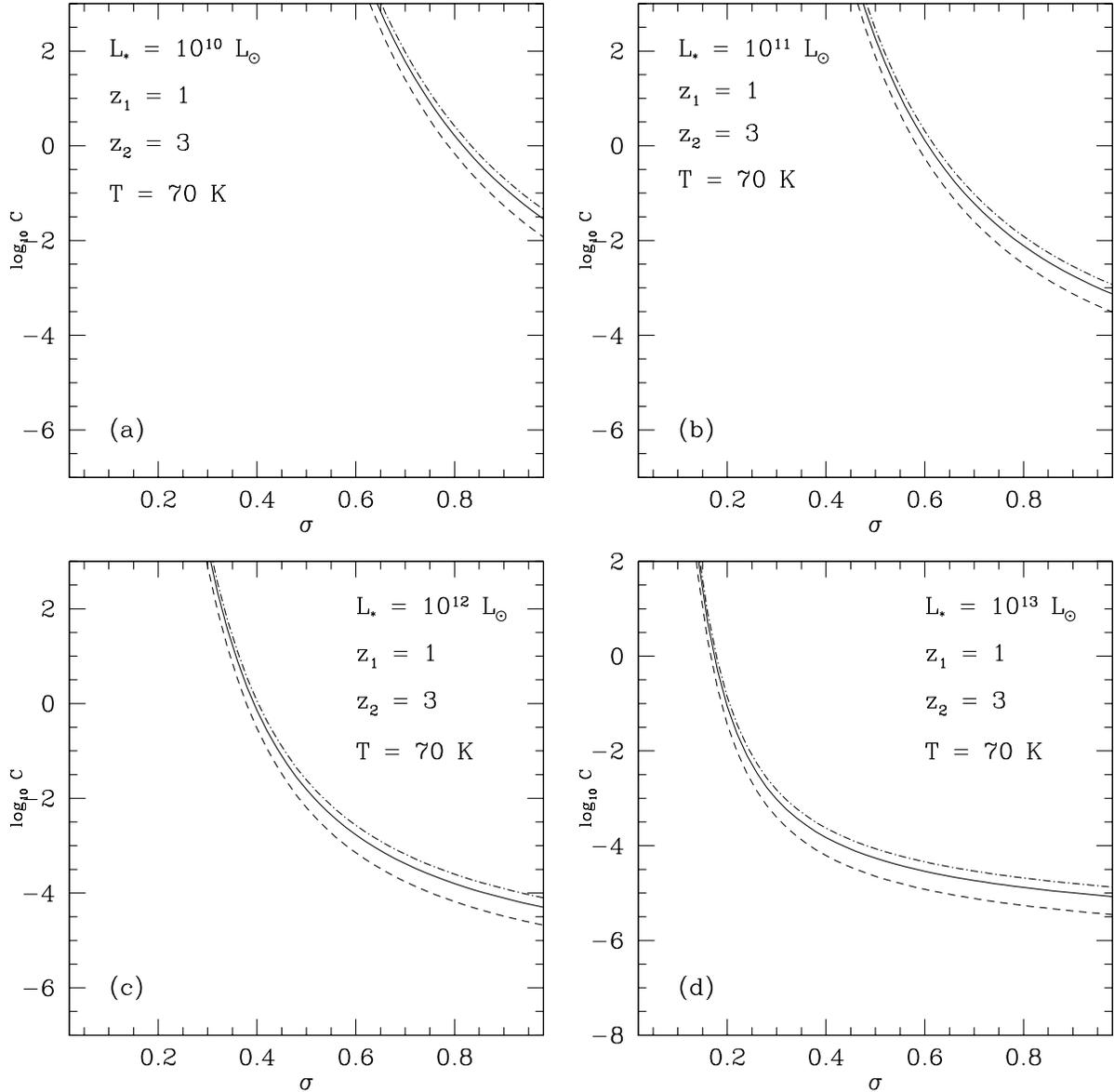, width=18.65cm}
\end{center}
{\vskip-7.2cm}
\caption{As Figure 2, but for $z_1=1$ and $z_2=3$.}
\end{minipage}
\end{figure*}

\begin{figure*}
\begin{minipage}{170mm}
{\vskip-3.5cm}
\begin{center}
\epsfig{file=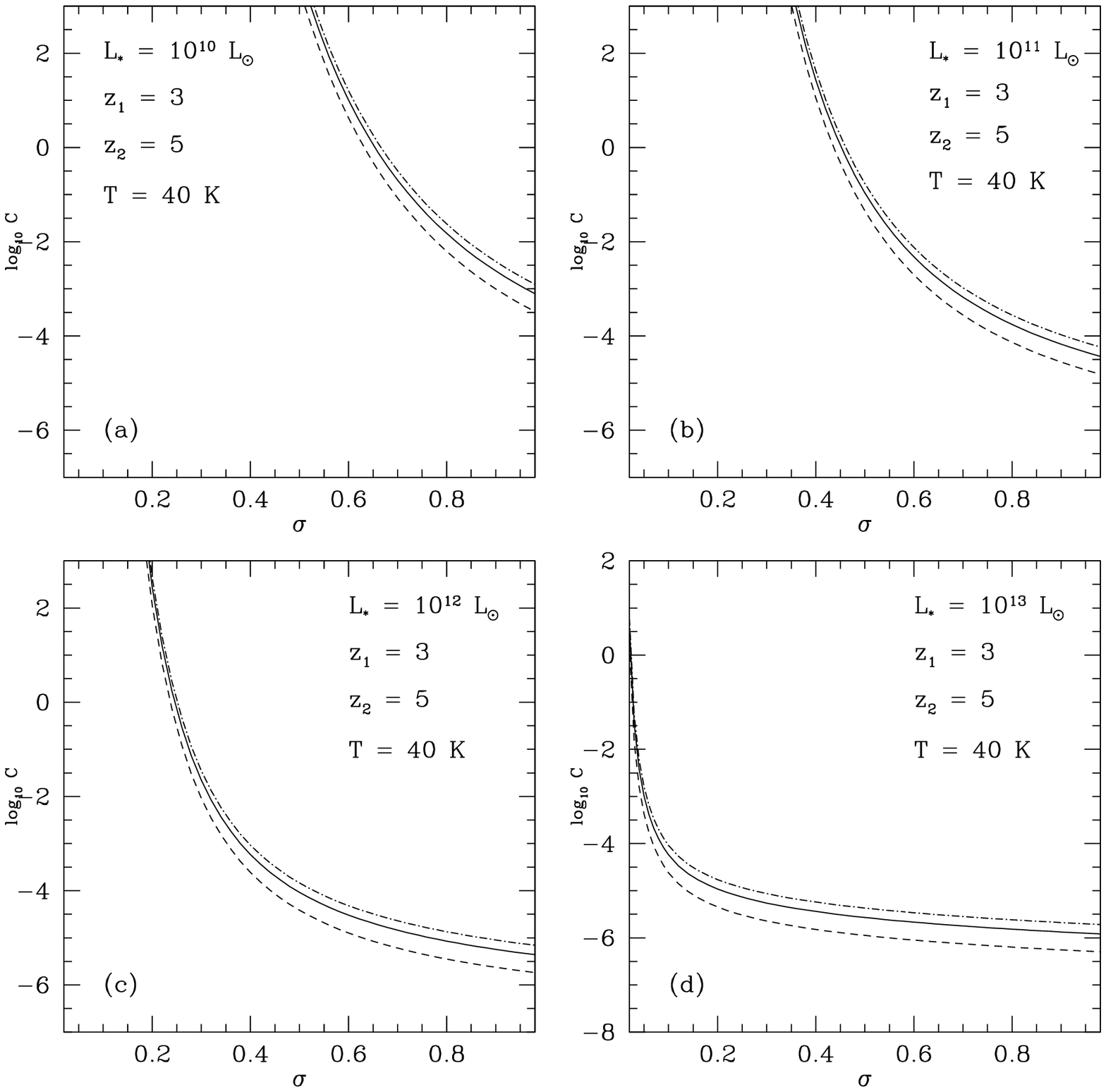, width=18.65cm}
\end{center}
{\vskip-7.2cm}
\caption{As Figure 2, but for $T=40$K.}
\end{minipage}
\end{figure*}

\begin{figure*}
\begin{minipage}{170mm}
{\vskip-3.5cm}
\begin{center}
\epsfig{file=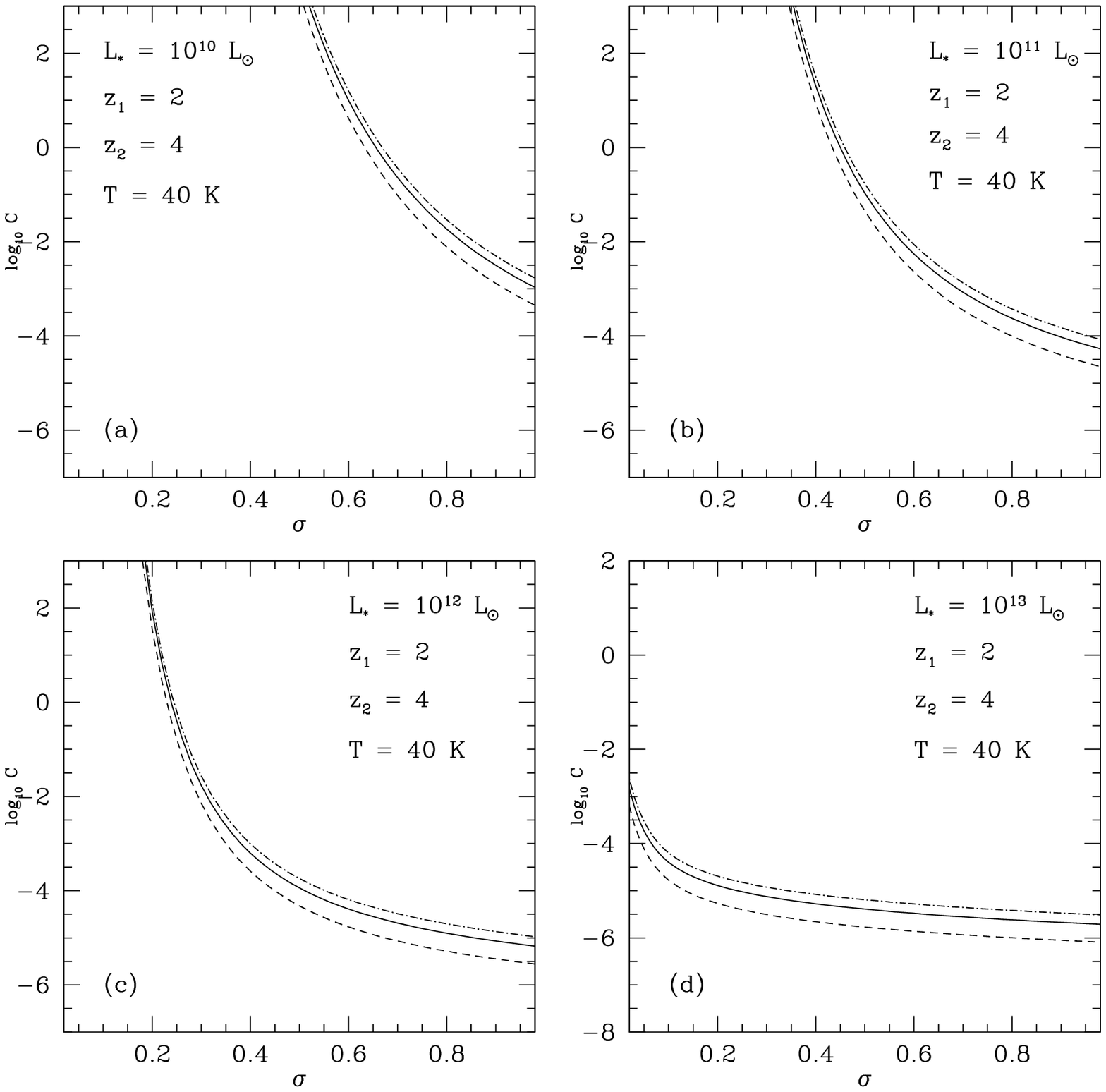, width=18.65cm}
\end{center}
{\vskip-7.2cm}
\caption{As Figure 3, but for $T=40$K.}
\end{minipage}
\end{figure*}

\begin{figure*}
\begin{minipage}{170mm}
{\vskip-3.5cm}
\begin{center}
\epsfig{file=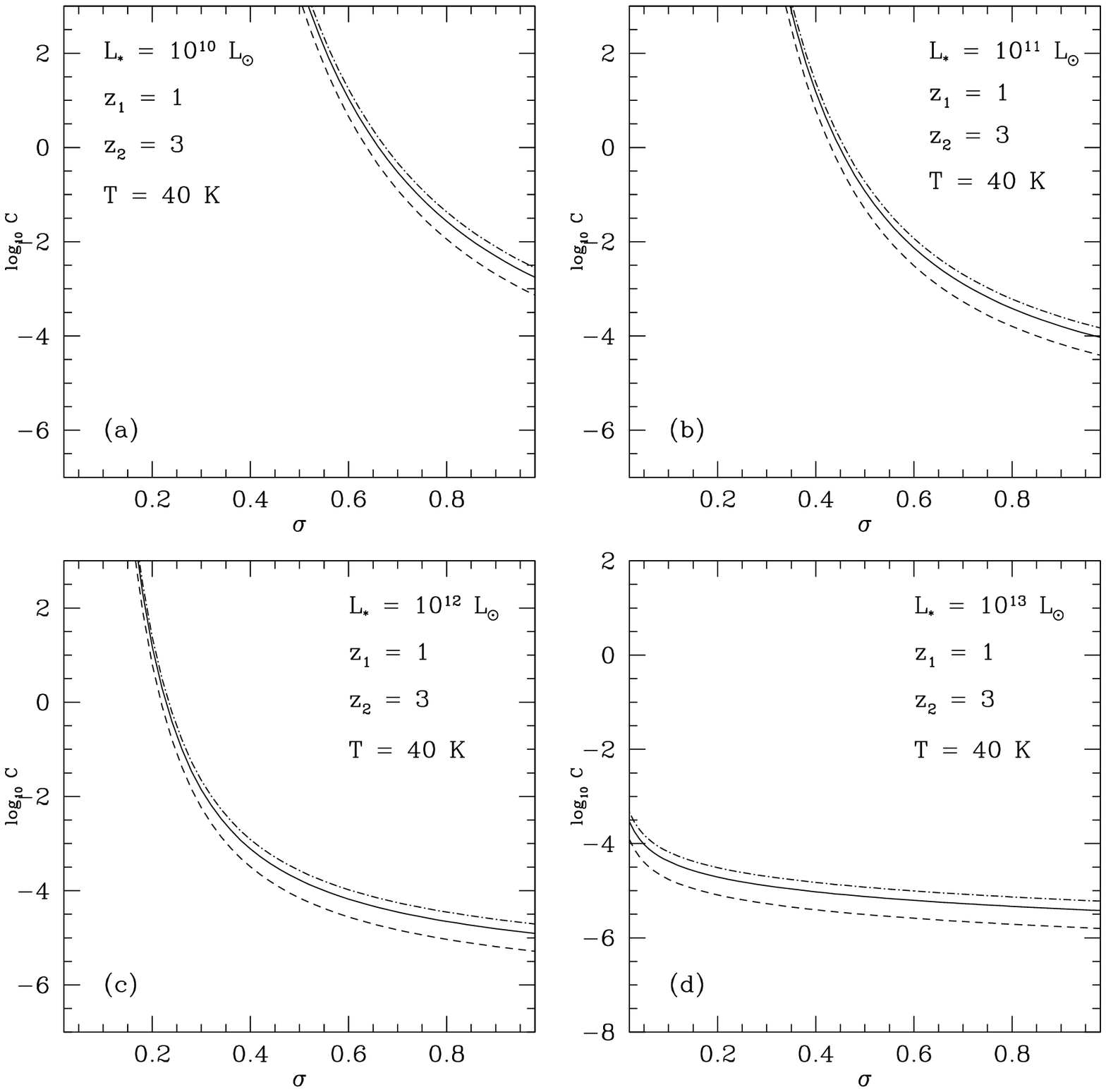, width=18.65cm}
\end{center}
{\vskip-7.2cm}
\caption{As Figure 4, but for $T=40$K.} 
\end{minipage}
\end{figure*} 

Higher temperature sources systematically produce greater background fluxes
and counts at shorter wavelengths, because their dust spectra peak at shorter 
wavelengths. Hence, lower normalizations are required in Fig.\,1 in order to 
explain the data for 40-K sources as compared with 70-K sources. The 40-K 
curve is consistent with a ratio of one over a range of luminosities, suggesting
that if a Saunders luminosity function describes the SCUBA sources, then they 
must have temperatures of about 40\,K. This was another conclusion of 
Blain et al.\ (1999a): see their Figure 4 -- in their favoured scenarios most of the 
SCUBA sources are at $z>2$ and have luminosities 
$\sim 10^{12} {\rm L}_{\odot}$ and dust temperatures $\sim 40$\,K. 

Finally, note that the results in Fig.\,1 are only weakly dependent on the
form of density evolution parameterized by $\gamma$.

\subsection{Comparison with the 850-$\mu$m source counts}

We now compute the normalizations $C$ required to fit the 850-$\mu$m source 
count data for a number of Gaussian luminosity functions (equation 1) with no 
density or luminosity evolution. $C$ is constant between redshifts 
$z_1$ and $z_2$ and is zero elsewhere. This is the simplest possible form of the 
luminosity function; at present the observations do not justify a more rigorous 
treatment. This parametrization is motivated in part by the
fact that the local 60-$\mu$m luminosity function (Saunders et al.\ 1990) is 
approximately Gaussian at the bright end. The local elliptical galaxy luminosity 
function (Binggeli, Sandage \& Tammann 1988, Ferguson \& Sandage 1991) is 
also Gaussian, which is relevant if the SCUBA sources evolve into elliptical 
galaxies: see Section 3. One important feature of the Gaussian function is that it 
is only valid for high-luminosity galaxies.  Large numbers of galaxies with low 
far-infrared luminosities will also exist between $z_1$ and $z_2$. Although these 
will not contribute to the source counts, they may contribute significantly to the 
infrared backgrounds if they are very numerous.  Hence, when we match our 
luminosity functions to counts (equation 5), we require the model prediction to
equal the observed values, but when we match our luminosity functions
to background fluxes (equation 3), we require the model prediction to be less 
than or equal to the observed values. Our approach is fundamentally different 
from that of Blain et al.\ (1999a). We are not evolving the local 60-$\mu$m 
luminosity function to high redshift.  Indeed, in Section 3, we argue that the
systems which have the highest far-infrared luminosities for redshifts 
$z_1 < z < z_2$ probably evolve into elliptical galaxies, 
which are not the most 
luminous systems at far-infrared wavelengths in the local Universe. 

In Figs\,2 to 7 we present the required normalization constants for
a Gaussian luminosity function to fit the 850-$\mu$m data of Smail
et al.\ (1997) for various values of $\sigma$ and $L_{*}$ given
a constant dust temperature $T$.  The uncertainties are large (see Table\,1),  
and so the range of acceptable normalizations is large, with a peak-to-peak 
spread of about a factor of 4. Given these large uncertainties, all the 
850-$\mu$m counts by different authors listed in
Table 1 are fully consistent.
Some general 
features of Figs\,2 to 7 are worth highlighting:
\begin{enumerate} 
\item $\sigma = 0$ (a $\delta$-function) never appears to fit the data well;  
\item lower normalizations are required for higher characteristic luminosities;
\item the absolute values of the normalizations depend primarily on where the 
dust spectrum peak is shifted in wavelength space. It is shifted closer to
850\,$\mu$m for a source at higher redshift at either 40 or 70\,K, and so a lower 
normalization is required at higher $z$;
\item lower temperatures require lower normalizations for a given 60-$\mu$m 
luminosity, since 850\,$\mu$m is closer to the blackbody peak. 
\end{enumerate} 

\subsection{Comparison with counts at other wavelengths}

Having isolated various models that are consistent with the 850-$\mu$m 
counts, we now consider the constraints from measurements at other
wavelengths. We select a number of models with various values of 
$\sigma$, $L_{*}$, $C$, and $T$, each represented by a point 
on the curves in Figs\,2 to 7, and thus consistent with the 850-$\mu$m count. 
For each model, we consider three separate scenarios: scenario ``0'', in which 
the 850-$\mu$m source counts are the mean Smail et al.\ (1997) values;
scenario ``+'', in which they are 1$\sigma$ larger; and scenario ``$-$'', in which 
they are 1$\sigma$ smaller.   The counts in the ``$-$'' scenario are very close to 
the lowest published counts (Barger et al.\ 1998a).  In all, a total of 45 models, 
each of which generates three scenarios, are considered.

In Table 2, we show how well the models defined by these parameters account 
for the measurements at other wavelengths.  The numbers in that table 
represent the model prediction relative to the observed counts or background 
fluxes listed in Table 1, assuming scenario ``0''. The fractional uncertainties in 
the numbers in Table 2 are given in Table 3. 

Other measurements exist in the literature that we do not use while
making this comparison.  For example, measurements of 15-$\mu$m source 
counts with {\it ISO} have recently been made (Rowan-Robinson 
et al.\ 1997, Aussel et al.~1998). 
However, these measurements probe far down the Wien tail of 
the dust spectrum, and are of very limited use in constraining the high-redshift 
luminosity function. Furthermore we would need to account for 
contamination in the 15-$\mu$m samples by nearby (unobscured) galaxies and 
AGN. Similarly the IRAS 60-$\mu$m source counts (Saunders et al.\ 1990)
are dominated by nearby ($z < 0.1$) sources. Measurements at 1.25\,mm offer a 
promising probe in the future, but to date measurements have only been made 
for IRAS-selected galaxies (Franceschini, Andreani \& Danese 1998). Unbiased 
samples are not yet available at this wavelength.

Some of the more notable general trends shown in Table 2 are:
\begin{enumerate}
\item models in which the sources are at higher redshift tend to
produce higher source counts or backgrounds at longer 
wavelengths, because the peak of the dust spectrum is shifted
to longer wavelengths;
\item models in which the sources are at higher redshift tend to
produce higher backgrounds for a given source count
{\it at the same wavelength}. The counts probe only part of the total range of 
galaxy luminosities, and there are a larger number
of low-luminosity sources that contribute to the backgrounds but
not to the counts;
\item on decreasing the temperature from 70 to 40\,K, the predicted 2.8-mm 
counts increase, but the predicted 175-$\mu$m counts and 140- and 
240-$\mu$m backgrounds decrease, because the peak in the dust emission 
spectrum shifts to longer wavelengths;
\item Changing the temperature has a substantial effect on the relative value 
of the source counts and backgrounds at the same wavelength. Decreasing the 
temperature has a greater relative effect on the counts, because the lower limit
in the integral in equation 5 is decreased: see Fig.\,8. 
\end{enumerate} 

Seven of the models that we investigate are consistent with all the
observations: see the last column of Table 2 for details. These all have a high 
characteristic luminosity greater than $10^{11} {\rm L}_{\odot}$.  
Using a different parametrization, Blain et al.\ (1999a) also found that most of the
SCUBA sources are distant galaxies with similarly high luminosities.
These seven models are:
12/40/3/5/0.9``0'' (hereafter Model A);
13/40/3/5/0.9``$-$'' (hereafter Model B);
11/40/2/4/0.9``$-$'' (hereafter Model C);
12/40/2/4/0.9``$-$'' (hereafter Model D);
13/40/2/4/0.5``0'' (hereafter Model E);
13/40/2/4/0.5``$-$'' (hereafter Model F);
13/40/1/3/0.1``$-$'' (hereafter Model G).
In general, the tables suggest that models with $T = 70$\,K overproduce the 
175-$\mu$m counts and the 140- and 240-$\mu$m backgrounds, given 
the normalization from the 850-$\mu$m source counts.  Models in which the 
galaxies are nearby, that is where $z_1 = 1$, tend to have similar problems. 
Conversely, models with very low temperatures $ T < 40$\, K would overproduce
the 2.8-mm counts, but such low-temperature sources are ruled out by the
consistency arguments of Blain et al.~(1999a --- see Section 4 of that paper).

\subsection{Comparison with the FIRAS background}

Fixsen et al.\ (1998) derived the cosmic submillimetre background from 
{\it COBE} FIRAS data.  Although the background at these wavelengths is 
dominated by emission from the Galaxy and the cosmic microwave background, 
they used three independent techniques to subtract out these signals. The 
resulting residual 850-$\mu$m extragalactic background radiation intensity 
$\nu I_\nu$=$0.55 \pm 0.15$\,nW\,m$^{-2}$\,sr$^{-1}$.   

We can predict 850-$\mu$m backgrounds in our models using
equation (3). The results are presented in Table 4.  As in the previous
section, we require that our models do not overproduce the background. Model 
G achieves this convincingly, and models B and F are consistent within 
2$\sigma$. 
We select these three models for further
study. In general, models with $z_1 = 3$ overpredict the 850-$\mu$m 
background if the 240-$\mu$m background is predicted correctly, because the 
observed spectra at a temperature of 40\,K are shifted to too long a 
wavelength. The surviving models have $z_1 = 2$ (B and F) or $z_1$ = 1 (G).
All these models also have $L_* = 10^{13}$\,L$_{\odot}$, meaning
that most of the far-infrared luminosity from this population comes
from galaxies with 60-$\mu$m luminosities in excess of 10$^{12}$\,L$_{\odot}$.  
As discussed in Section\,2, models with lower characteristic luminosities tend 
to overpredict the 850-$\mu$m background if they match the 850-$\mu$m 
counts. 

Note that in the three surviving models (B, F and G) the normalization of the 
850-$\mu$m counts is consistent with all the observations. In the best-fitting 
model (G), the mean redshift of the sources $z_{\rm s} \sim 2$, as argued by Lilly 
et al.\ (1998). One half (F,G) or more (B) of the 240-$\mu$m background, but 
considerably less of the 140-$\mu$m background, comes from the 
high-luminosity galaxies described by our Gaussian luminosity function. In 
models B and F, about half of the 175-$\mu$m sources are the same sources 
that contribute to the 850-$\mu$m source counts; in model G, the two 
populations are practically the same. In Model B, the predicted 2.8-mm 
count is close to the observed limit (Wilner \& Wright 1997); for the other models
the predicted count is much smaller. The 450-$\mu$m source count limit
(Barger et al.\ 1998a) is larger than the predicted values by a factor of about 3 
for Model B, and by a much larger factor for the other models.

Three models appear to fit all the data. Even within our simple parametrization 
we do not find a unique best-fitting model. Furthermore, there are presumably 
many other (non-Gaussian) models that fit all the data, for example the 
models of Blain et al.\ (1999a).  Nevertheless most of our models seem to be ruled 
out when we consider all the observations in conjunction. That a relatively 
narrow region of parameter space is consistent with observation 
($L_{*} \sim 10^{13}$\,L$_{\odot}$, $T \simeq 40$\,K, $z_{\rm s} \sim 2$) is 
encouraging and suggests that it will be productive to investigate these models 
in further detail. 

\section{Properties of the SCUBA sources} 

We have isolated three models of the redshift-dependent luminosity
function of ULIRGs that are consistent with observation, and now
investigate the cosmological implications. 
In Section 3.1 we investigate the possibility that these are dusty star-forming
galaxies. We use the star-formation models of Leitherer \& Heckman (1995) to
place the galaxies on the Madau Plot, and then examine the consequences
of this interpretation in the context of the wider galaxy formation
picture. In Section 3.2 we investigate the alternative possibility that the 
sources are dust-enshrouded AGNs. 

\subsection{The SCUBA sources as star-forming galaxies}

The majority of local ULIRGs appear to be star-forming galaxies and not 
dust-enshrouded AGN (Sanders et al.\ 1988a, Genzel et al.\ 1988). Furthermore, 
the temperatures of the SCUBA sources inferred in Section 2 ($T \sim 40$\,K) 
are close to, although systematically slightly cooler than, those of local 
star-forming ULIRGs.  Locally, dust-enshrouded AGN like Mrk 231 tend to have 
higher dust temperatures than dust-enshrouded starbursts like Arp 220 
(Sanders et at.\ 1988a); however, note that some dusty high-redshift quasars 
appear to be fairly cold: see Benford et al.\ (1998). Hence, it seems 
reasonable to investigate the possibility that the SCUBA sources are
high-redshift star-forming galaxies.

\subsubsection{Leitherer-Heckman models}

In order to compute the cosmological star-formation rate associated with the 
SCUBA sources at high redshift, we need a recipe to convert far-infrared 
luminosities to star-formation rates for individual sources.

The transformation between these two quantities is not straightforward to 
determine observationally even for local ULIRGs because of the high internal 
extinction along our lines to the galaxy centers. For example, 
even the near-infrared 2.2-$\mu$m Br$\gamma$ line strength--far-infrared
luminosity correlation (Goldader et al.\ 1997a,b), which is usually very 
reliable, breaks down in ULIRGs at very high luminosities, presumably due to 
internal extinction. We therefore need to compute this transformation using 
models (Leitherer \& Heckman 1995).

\begin{table*} 
\caption{Comparison between observations and models. All the numbers 
represent the predicted source counts or background flux for
model ``0'' that are appropriate to the luminosity function defined by
the parameters $L_{*}$, $T$, $\sigma$, $z_1$, and $z_2$, relative to the 
observed counts and background intensities listed in Table\,1.
The model parameters are listed in condensed form in the model name, which
takes the general form log$_{10}$($L^*_{60}$)/$T$/$z_1$/$z_2$/$\sigma$. The 
figure panel in which the constraints for each model are imposed are 
listed in the second column.  
A value less than $10^{-10}$ is listed as $\sim 0$ in the Table.
The errors in these quantities are listed in Table\,3. 
All the models have been selected {\it a priori} to give the correct 
850-$\mu$m counts. The comments refer to the selection or rejection of 
models based on the listed values. The range of acceptable values and a 
description of the numerical code values are given in Table\,3. The 
140-$\mu$m and 240-$\mu$m background values of 
Schlegel et al.\ (1998), the 175-$\mu$m counts of Lagache
et al.~(1998), and the 450-$\mu$m counts of Smail et al.~(1997)
are fitted. 
} 
{\vskip -2.75mm}
{$$\vbox{
\halign {\hfil #\hfil && \quad \hfil #\hfil \cr
\noalign{\hrule \medskip}
Model & Figure 
         & 2.8-mm & 450-$\mu$m & 175-$\mu$m
         & 240-$\mu$m & 140-$\mu$m & comments &\cr
       &  & counts & counts & counts & background & background & (See
Table\,3)&\cr
\noalign{\smallskip \hrule \smallskip}
\cr

10/70/3/5/0.5 & 2a & $\sim 0$
        & $7.2 \times 10^{-5}$ 
        & $0.18$  
        & $1.5 \times 10^{10}$ & $7.5 \times 10^9$ & (4$^{-}$56) &\cr  
10.70/3/5/0.9 & 2a & $0.0019$
        & $0.082$
        & $2.5$  
        & $110$ & $55$ & (3$^{0+}$56) &\cr
11/70/3/5/0.5 & 2b & $3.2 \times 10^{-9}$
        & $8.1 \times 10^{-4}$
        & $0.40$  
        & $2.3 \times 10^{5}$ & $1.1 \times 10^{5}$ & (56) &\cr
11/70/3/5/0.9 & 2b & $0.011$
        & $0.17$
        & $3.9$ 
        & $15$ & $7.4$ & (356) &\cr
12/70/3/5/0.5 & 2c & $1.3 \times 10^{-6}$
        & 0.0090 
        & $1.0$ 
        & $178$ & $87$ & (3$^{+}$56) &\cr
12/70/3/5/0.9 & 2c & $0.067$
        & $0.36$
        & $6.0$ 
        & $6.3$ & $3.1$ & (356$^{0+}$) &\cr
13/70/3/5/0.1 & 2d & $\sim 0$
        & $\sim 0$
        & $3.2 \times 10^{-6}$ 
        & $2.9 \times 10^{13}$ & $1.9 \times 10^{13}$ & (456) &\cr
13/70/3/5/0.5 & 2d & $4.4 \times 10^{-4}$
        & $0.094$
        & $3.0$ 
        & $5.8$ & $2.8$ & (356$^{0+}$) &\cr
13/70/3/5/0.9 & 2d & $0.35$
        & $0.71$
        & $9.6$ 
        & $7.2$ & $3.5$ & (2$^{+}$356) &\cr
10/70/2/4/0.5 & 3a & $\sim 0$
        & $2.3 \times 10^{-4}$
        & $32$ 
        & $1.9 \times 10^{10}$ & $1.6 \times 10^{10}$ & (356) &\cr
10/70/2/4/0.9 & 3a & $0.0014$
        & $0.12$
        & $14$  
        & $150$ & $120$ & (356) &\cr
11/70/2/4/0.5 & 3b & $1.5 \times 10^{-9}$
        & $0.0021$
        & $20$ 
        & $3.0 \times 10^{5}$ & $2.5 \times 10^{5}$ & (356) &\cr
11/70/2/4/0.9 & 3b & $0.0090$
        & $0.23$
        & $14$ 
        & $20$ & $17$ & (356) &\cr
12/70/2/4/0.5 & 3c & $7.2 \times 10^{-7}$
        & $0.017$
        & $14$ 
        & $230$ & $190$ & (356) &\cr
12/70/2/4/0.9 & 3c & $0.055$ 
        & $0.45$
        & $16$ 
        & $8.3$ & $7.0$ & (356) &\cr
13/70/2/4/0.1 & 3d & $\sim 0$
        & $\sim 0$
        & $5.6 \times 10^{5}$ 
        & $2.4 \times 10^{13}$ & $2.1 \times 10^{13}$ & (356) &\cr
13/70/2/4/0.5 & 3d & $3.0 \times 10^{-4}$
        & $0.15$
        & $14$ 
        & $7.5$ & $6.4$ & (356) &\cr
13/70/2/4/0.9 & 3d & $0.3$
        & $0.82$
        & $18$ 
        & $9.5$ & $8.0$ & (356) &\cr
10/70/1/3/0.5 & 4a & $ \sim 0$
        & $8.8 \times 10^{-4}$
        & $3600$ 
        & $1.7 \times 10^{10}$ & $2.4 \times 10^{10}$ & (356) &\cr
10/70/1/3/0.9 & 4a & $0.0011$
        & $0.17$
        & $60$ 
        & $170$ & $230$ & (356) &\cr
11/70/1/3/0.5 & 4b & $7.6 \times 10^{10}$
        & $0.0054$
        & $560$ 
        & $3.0 \times 10^5$ & $4.0 \times 10^5$ & (356) &\cr
11/70/1/3/0.9 & 4b & $0.0073$
        & $0.31$
        & $44$ 
        & $23$ & $32$ & (356) &\cr
12/70/1/3/0.5 & 4c & $4.5 \times 10^{-7}$
        & $0.034$
        & $120$ 
        & $250$ & $350$ & (356) &\cr
12/70/1/3/0.9 & 4c & $0.047$
        & $0.55$
        & $34$ 
        & $10$ & $14$ & (356) &\cr
13/70/1/3/0.1 & 4d & $ \sim 0 $
        & $ \sim 0$ 
        & $3.1 \times 10^{10}$ 
        & $3.0 \times 10^{11}$ & $4.3 \times 10^{11}$ & (356) &\cr
13/70/1/3/0.5 & 4d & $2.2 \times 10^{-4}$
        & $0.21$
        & $41$ 
        & $8.8$ & $12$ & (356) &\cr
13/70/1/3/0.9 & 4d & $0.27$
        & $0.94$
        & $29$ 
        & $12$ & $16$ & (2$^{+}$356) &\cr
10/40/3/5/0.5 & 5a & $1.0 \times 10^{-8}$             
        & $9.2 \times 10^{-6}$
        & $1.1 \times 10^{-8}$ 
        & $1.1 \times 10^{6}$ & $7.1 \times 10^{5}$ & (456) &\cr 
10/40/3/5/0.9 & 5a & $0.018$
        & $0.033$
        & $0.0067$ 
        & $3.6$ & $0.23$ & (45) &\cr
11/40/3/5/0.5 & 5b & $2.1 \times 10^{-6}$
        & $2.1 \times 10^{-4}$
        & $1.4 \times 10^{-6}$ 
        & $200$ & $14$ & (456) &\cr
11/40/3/5/0.9 & 5b & $0.087$
        & $0.092$
        & $0.034$ 
        & $0.99$ & $0.067$ & (45$^{+}$) &\cr
12/40/3/5/0.5 & 5c & $3.9 \times 10^{-4}$
        & $0.0050$
        & $1.8 \times 10^{-4}$ 
        & $1.7$ & $0.12$ & (45$^{0+}$) &\cr
12/40/3/5/0.9 & 5c & $0.40$
        & $0.25$
        & $0.17$ 
        & $0.84$ & $0.056$ & (4$^{-}$5$^{+}$) &\cr
13/40/3/5/0.1 & 5d & $\sim 0$
        & $3.3 \times 10^{-18}$
        & $\sim 0$ 
        & $1.14$ & $0.076$ & (45$^{+}$) &\cr
13/40/3/5/0.5 & 5d & $0.053$
        & $0.11$
        & $0.019$ 
        & $0.50$ & $0.034$ & (4) &\cr
13/40/3/5/0.9 & 5d & $1.6$
        & $0.63$
        & $0.82$ 
        & $1.8$ & $0.12$ & (1$^{0+}$3$^{+}$5$^{0+}$) &\cr
10/40/2/4/0.5 & 6a & $2.96 \times 10^{-9}$
        & $8.6 \times 10^{-5}$
        & $5.7 \times 10^{-4}$ 
        & $1.1 \times 10^{6}$ & $2.1 \times 10^{5}$ & (456) &\cr
10/40/2/4/0.9 & 6a & $0.012$
        & $0.068$
        & $0.22$ 
        & $6.6$ & $1.2$ & (56$^{+}$) &\cr
11/40/2/4/0.5 & 6b & $8.7 \times 10^{-7}$
        & $0.0012$
        & $0.0058$ 
        & $280$ & $51$ & (4$^{0-}$56) &\cr
11/40/2/4/0.9 & 6b & $0.065$
        & $0.16$
        & $0.54$ 
        & $2.0$ & $0.36$ & (5$^{0+}$) &\cr
12/40/2/4/0.5 & 6c & $2.2 \times 10^{-4}$
        & $0.017$
        & $0.062$ 
        & $3.0$ & $0.54$ & (4$^{0-}$5$^{0+}$) &\cr
12/40/2/4/0.9 & 6c & $0.33$
        & $0.38$
        & $1.4$ 
        & $1.77$ & $0.32$ & (3$^{0+}$5$^{0+}$) &\cr
13/40/2/4/0.1 & 6d & $\sim 0$
        & $\sim 0$
        & $2.0 \times 10^{-9}$ 
        & $1.1$ & $0.020$ & (45$^{+}$) &\cr
13/40/2/4/0.5 & 6d & $0.040$
        & $0.22$
        & $0.70$ 
        & $1.1$ & $0.19$ & (5$^{+}$) &\cr
13/40/2/4/0.9 & 6d & $1.4$
        & $0.84$
        & $3.7$ 
        & $4.0$ & $0.72$ & (1$^{0+}$2$^{+}$35) &\cr
10/40/1/3/0.5 & 7a & $9.4 \times 10^{-10}$
        & $8.0 \times 10^{-4}$
        & $3.7$ 
        & $9.6 \times 10^{5}$ & $4.3 \times 10^{5}$ & (356) &\cr
10/40/1/3/0.9 & 7a & $0.0086$
        & $0.13$
        & $3.6$ 
        & $11$ & $4.8$ & (356) &\cr
11/40/1/3/0.5 & 7b & $4.0 \times 10^{-7}$
        & $0.0062$
        & $3.0$ 
        & $323$ & $147$ & (356) &\cr
11/40/1/3/0.5 & 7b & $0.051$
        & $0.28$
        & $4.5$ 
        & $3.5$ & $1.6$ & (356$^{0+}$) &\cr
12/40/1/3/0.1 & 7c & $ \sim 0$
        & $\sim 0$
        & $140$ 
        & $2.8 \times 10^{17}$ & $1.3 \times 10^{17}$ & (356) &\cr
12/40/1/3/0.5 & 7c & $1.4 \times 10^{-4}$
        & $0.52$
        & $3.0$ 
        & $4.6$ & $2.1$ & (356$^{0+}$) &\cr
12/40/1/3/0.9 & 7c & $0.28$
        & $0.56$
        & $6.5$ 
        & $3.3$ & $1.5$ & (356$^{0+}$) &\cr
13/40/1/3/0.1 & 7d & $\sim 0$
        & $6.1 \times 10^{-6}$
        & $2.2$ 
        & $1.2$ & $0.54$ & (3$^{0+}$5$^{+}$) &\cr
13/40/1/3/0.5 & 7d & $0.033$
        & $0.42$
        & $5.4$ 
        & $2.0$ & $0.92$ & (35$^{0+}$6$^{+}$) &\cr
13/40/1/3/0.9 & 7d & $1.3$
        & $1.1$
        & $10$ 
        & $8.0$ & $3.6$ & (356) &\cr
\noalign{\smallskip \hrule}
\noalign{\smallskip}\cr}}$$}
\end{table*}

\begin{table*}
\caption{Fractional uncertainties in the background and source counts 
values used to accept or reject the models listed in Table\,2. The acceptable 
ranges of values are used to make the selection, and the relevant cases are 
described by numerical codes in the last column of Table\,2. A code is 
shown if the model is rejected on the grounds of: (1) overproduction of 
the 2.8-mm count; (2) overprediction of the 450-$\mu$m count; (3) 
overproduction of the 175-$\mu$m count; (4) underprediction of the 
175-$\mu$m count; (5) overprediction of the {\it COBE} 240-$\mu$m background; 
and (6) overprediction of the {\it COBE} 140-$\mu$m background. Although 
a model that underpredicts the 175-$\mu$m count is not formally excluded, 
such a model requires that the 175-$\mu$m counts and 850-$\mu$m counts come
from entirely different populations of galaxies. This is not impossible, but we 
regard it as unlikely; we choose a factor of ten discrepancy as the cutoff
between a model being acceptable and unacceptable.   
In the last column of Table\,2, superscripts 0, + or $-$ indicate that 
the exclusion applies only in that scenario, corresponding to the 
values listed below. The absence of a superscript implies that the 
model is rejected in all three scenarios.} 
{\vskip -0.75mm}
{$$\vbox{
\halign {\hfil #\hfil && \quad \hfil #\hfil \cr
\noalign{\hrule \medskip}
Model & 2.8-mm
          & 450-$\mu$m
         & 175-$\mu$m  & 240-$\mu$m & 140-$\mu$m &\cr
       & counts & counts & counts & background & background & &\cr
\noalign{\smallskip \hrule \smallskip}
\cr
Fractional Error in ratio & $-$ & $-$ & 0.15 & 0.24 & 0.41 &\cr
 & & & & & & & &\cr
Range of Acceptability (``0'' models) &
$<1.0$ & $<1.0$ &   [0.85,1.2] &  $<1.2$ & $<1.4$ &\cr
Range of Acceptability (``$-$'' models) &
$<2.4$ & $<2.4$ &   [2.0,2.8] &  $<3.0$ & $<3.4$ &\cr
Range of Acceptability (``+'' models) &
$<0.63$ & $<0.63$ & [0.54,0.73] &  $<0.78$ & $<0.89$ &\cr
 & & & & & & & &\cr
\noalign{\smallskip \hrule}
\noalign{\smallskip}\cr}}$$}
\end{table*}

In Table 5 we list a number of star-formation models that we investigate in 
further detail. Note that we use capital letters to represent galaxy formation 
models (Section 2) and roman numerals to represent the different star-formation 
models. There is insufficient information to determine accurately the appropriate 
set of input parameters to the star-formation models for the SCUBA sources, 
and even for local ULIRGs in many cases. The more important parameters are:
\begin{enumerate}
\item the mode of star formation.  For example, the star formation may occur
continuously, as an instantaneous burst, or with a more complex time
dependence, for example an exponential decay. We investigate both the 
instantaneous (Models I and II) and continuous (models III through X) cases.
Observations of the core of Arp 220 (Mouri \& Taniguchi 1992;
Prestwich, Joseph, \& Wright 1994; Armus et al.\ 1995; Larkin et al.\ 1995,
Scoville et al.\ 1998), combined with the inferred presence of ionizing sources 
there (Larkin et al.\ 1995; Kim et al.\ 1995; Goldader et al.\ 1995), indicate that 
the continuous models offer a more plausible description of the star formation 
in this well studied ULIRG;
\item the stellar initial mass function (IMF). The precise form of the IMF in 
starburst galaxies is very much an open question: see Leitherer (1998) for a 
review. It appears that the IMF of the most luminous star clusters in the Milky 
Way and the Magellanic Clouds follows closely the Salpeter form (Hunter et al.\ 
1995; Hunter et al.\ 1997; Massey et al.\ 1995a). However, the field-star IMF may 
be significantly steeper than the cluster IMF (Massey et al.\ 1995b),
and there is some evidence (Meurer et al.\ 1995) that the most massive stars
in starbursts form in environments that are more similar to the field than to 
the cores of the luminous star clusters studied above. There are 
hence few direct observations to guide us to the correct IMF in starburst 
galaxies, and we therefore consider four different scenarios in Models III to X.
Leitherer \& Heckman (1995) consider power-law IMFs, and 
we investigate various possible combinations of
the upper and lower mass cutoffs and the slope of this power law.
Note that if the SCUBA sources end up as elliptical galaxies (Section 3.1.3; 
Blain et al. 1999a; Lilly et al.\ 1998), a lower limit to the IMF at 
$M_{\rm l}$ = 3\,M${_\odot}$ is suggested, in agreement with the value suggested 
by Zepf \& Silk (1996).
\item the inital gas metallicity.  We assume a metallicity of twice solar.
This is close to the metallicities of the most luminous elliptical
galaxies (e.g.~Faber 1973, Vader 1986); 
\item the age at which we observe the star clusters.  We investigate the
cases where we observe the galaxies at between 10$^8$ and 10$^{8.5}$
years after the onset of star formation.  Fortunately, most of the
properties of the starburst, like the total bolometric luminosity, reach
a plateau soon after the onset of star formation, and so our results do not 
depend strongly on this parameter.  The median total amount
of gas consumed by a $L_{*}$ galaxy 10$^{8.5}$ years after the onset
of star-formation is 7 $\times 10^{10}$\,M$_\odot$ in our models.  This is 
consistent with the progenitor galaxies to the SCUBA sources being gas-rich 
normal galaxies; 
\item the effects of extinction.  Since we are only attempting to model
the far-infrared and submillimetre properties of the starburst, we
can assume that the galaxies are optically thin and do not need to
make any corrections for extinction. We assume that the dust absorbs
and reradiates all the energy produced by the starburst.
\end{enumerate} 

From the seventh column of Table 5 it is immediately apparent that there is 
more than an order of magnitude of uncertainty in the transformation from 
far-infrared luminosity to star-formation rate, due to the uncertainty in the
star-formation parameters.  The constant conversion factor of 2.2 $\times$
10$^9$\,L$_{\odot}$\,M$_{\odot}^{-1}$\,yr (Rowan-Robinson et al.\ 1997), 
used by Blain et al.\ (1999a), corresponding to 0.22 in the seventh column of
Table 5, is bracketed by our results.

\subsubsection{The Madau plot}

We combine our redshift-dependent luminosity functions (Models B, F and G) 
with the star-formation parameters for the models listed in Table 5 
(Models I to X) to compute the comoving star formation rate in the SCUBA 
sources. The results are listed in Table 6.  We can then put the SCUBA 
sources on the the Madau plot: see Figs\,9, 10 and 11. The other points on the 
Madau plot all come from optically-selected samples. As there is no luminosity 
evolution in the models between $z_1$ and $z_2$, the comoving star-formation 
rate is constant between these redshifts, and so the lines on the Madau plot are 
horizontal. 

The figures show that the uncertainty in determining where the SCUBA 
sources lie on the Madau plot is huge, more than an order of magnitude, 
even if we have specified the correct luminosity--redshift distribution model B, 
F or G. This uncertainty is solely due to our lack of knowledge regarding the 
star-formation parameters, as discussed in Section 3.1.1. For Model B, and 
possibly Model F, if the true cosmic star-formation rate in the SCUBA sources
is towards the higher end of our permitted range, then they dominate the 
star-formation rate of the Universe: this is essentially the scenario proposed
by Blain et al.\ (1999a) and Hughes et al.\ (1998). If the true cosmic star-formation 
rate in the SCUBA sources is towards the lower end of our range, then
they do not contribute significantly to the star-formation rate of the Universe 
at any redshift. In general, the instantaneous models 
predict higher star-formation rates.  This is because more stars
need to be formed to produce a given bolometric luminosity we observe
more than 10$^8$ years after the starburst.

It is interesting to note that the IMF proposed by Zepf \& Silk (1996) for 
elliptical galaxies, if valid for the SCUBA sources, results in them being at 
the extreme low end of our proposed range.  This is because these models 
underproduce low-mass stars, and so result in a low star-formation rate for 
a given amount of bolometric luminosity: the low-mass stars, when young, 
do not contribute significantly to the bolometric luminosity. The implication 
here is that if SCUBA sources evolve into elliptical galaxies, then they do 
not contribute significantly to the star-formation rate of the Universe at 
any redshift given our models of the luminosity function. 
 
\begin{figure}
\begin{center}
\epsfig{file=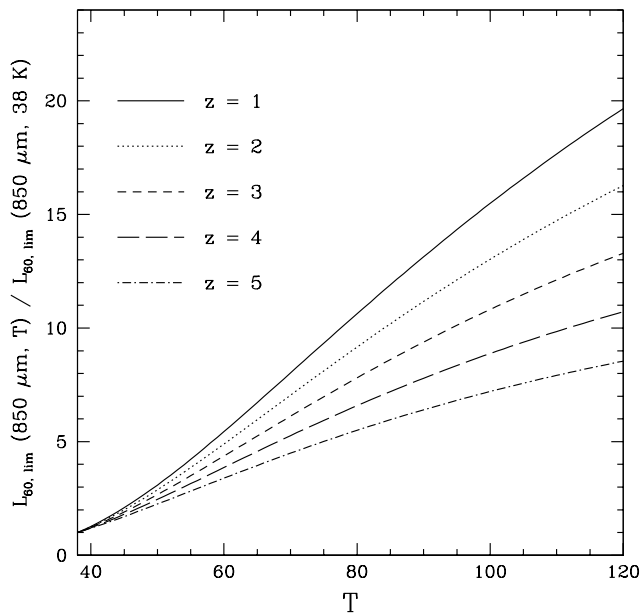, width=8.65cm}
\end{center}
\caption{The lower limit on the integral in equation (5) as a function of the 
galaxy dust temperature $T$, normalized to its value for $T=38$\,K, predicted 
from {\it IRAS} and {\it ISO} counts by Blain et al.\ (1999a).} 
\end{figure} 

The broadband optical colours of the SCUBA sources are not significantly 
different from those of normal field galaxies (Smail et al.\ 1998). Furthermore, 
two of the three local ULIRGs studied at ultraviolet wavelengths with {\it HST} 
by Trentham, Kormendy \& Sanders (1998) would also have broadband colours 
similar to normal field galaxes if they were placed at high redshift. Therefore,
SCUBA sources cannot be identified as submillimetre-luminous galaxies 
based on broadband optical colours alone. The star-formation rates given 
in Table 5 are far higher than the rates in normal galaxies. Therefore we need 
to treat the SCUBA sources as a separate population on the Madau plot. This 
is particularly important if the SCUBA sources contribute significantly to the
star-formation rate of the Universe at any redshift. 

\begin{table}
\caption{Predicted 850-$\mu$m background intensities. The observed
850-$\mu$m background intensity is $0.55 \pm 0.15$\,nW\,m$^{-2}$\,sr$^{-1}$ 
(Fixsen et al. 1998).} 
{$$\vbox{
\halign {\hfil #\hfil && \quad \hfil #\hfil \cr
\noalign{\hrule \medskip}
Model  & Background / & Model  & Background /\cr
       & nW\,m$^{-2}$\,sr$^{-1}$  &  & nW\,m$^{-2}$\,sr$^{-1}$\cr
\noalign{\smallskip \hrule \smallskip}
\cr
A & 3.40 & E & 1.71 \cr
B & 0.85 & F & 0.71 \cr
C & 1.32 & G & 0.34 \cr
D & 1.19 & & \cr
\noalign{\smallskip \hrule}
\noalign{\smallskip}\cr}}$$}
\end{table}

\begin{table*}
\caption{Properties of star-formation models. 
$\alpha$ is the slope of the stellar IMF and $M_{\rm u}$ and  $M_{\rm l}$
are the lower and upper mass cutoffs.
The final two columns are derived from the results of Leitherer \& Heckman
(1995).  The star-formation
rates (SFR) as a function of 60-$\mu$m luminosity come from Figs\,7 and 8,
assuming a temperature of 70\,K.  The metal masses $M_Z$ come from 
Figs\,53 and 54 of Leitherer \& Heckman (1995), assuming that the mass of 
metals produced is equal to the mass returned by winds and supernovae.
All models assume an initial metallicity of twice solar.
} 
{$$\vbox{
\halign {\hfil #\hfil && \quad \hfil #\hfil \cr
\noalign{\hrule \medskip}
Model & SF profile & log$_{10}$(age/yr) &
      $\alpha$ & $M_{\rm u}$ / M$_\odot$ & $M_{\rm l}$ / M$_\odot$ & 
SFR / ($L_{60} / 10^{9} {\rm L}_{\odot}$) M$_\odot$\,yr$^{-1}$ 
& $M_Z$ / ($L_{60} / 10^{9} {\rm L}_{\odot}$) M$_\odot$ &\cr
\noalign{\smallskip \hrule \smallskip}
\cr
SF I    & Instantaneous & 8.0 & 2.35 & 100 & 1 & 1.0 & $4.0 \times 10^7$ &\cr
SF II   & Instantaneous & 8.5 & 2.35 & 100 & 1 & 0.81 & $1.0 \times 10^8$ &\cr
SF III  & Continuous    & 8.0 & 2.35 & 100 & 1 & 0.11 & $3.5 \times 10^6$ &\cr
SF IV   & Continuous    & 8.5 & 2.35 & 100 & 1 & 0.072 & $7.2 \times 10^6$ &\cr
SF V    & Continuous    & 8.0 & 2.35 &  30 & 1 & 0.15 & $4.1 \times 10^6$ &\cr
SF VI   & Continuous    & 8.5 & 2.35 &  30 & 1 & 0.12 & $1.0 \times 10^7$ &\cr
SF VII  & Continuous    & 8.0 & 3.3  & 100 & 1 & 0.34 & $8.0 \times 10^5$ &\cr
SF VIII & Continuous    & 8.5 & 3.3  & 100 & 1 & 0.25 & $1.6 \times 10^6$ &\cr
SF IX   & Continuous    & 8.0 & 2.35 & 100 & 3 & 0.056 & $3.5 \times 10^6$ &\cr
SF X    & Continuous    & 8.5 & 2.35 & 100 & 3 & 0.043 & $7.2 \times 10^6$ &\cr
\noalign{\medskip \hrule}
\noalign{\smallskip}\cr}}$$}
\end{table*}

\subsubsection{Why do the result differ from those of other authors} 

Both Blain et al.~(1999a) and Hughes et al.~(1998) derive global star-formation rates
at least an order of magnitude higher than we derive in the current paper.
There are two sources of discrepancy.
\vskip 1pt
\noindent (i) differences in star-formation rates. 
For example Blain et al.~(1999a) use a value of $M_{\rm l}$ below 1 M$_\odot$ and
consequently assume star-formation rates that are towards the upper end of our
considered range.
\vskip 1pt
\noindent (ii) differences in the assumed luminosity-redshift distribution, which can 
be broadly split further into (a) differences in the characteristic luminosity of the
sources who dominate the contribution to the SCUBA counts , and (b) difference in the
normalization of the luminosity function at a given redshift.  Effect (iia) is not a
major source of discrepancy between our results and those of Blain et al.~(1999a) --- in
both cases, most of the star formation occurs in very luminous sources
with $L_{60} > 10^{11}$\,L$_{\odot}$.  Effect (iib) is a much more substantial source
of discrepancy.  Blain et al.~(1999a) assume a redshift-dependent Saunders luminosity function at
high redshift which is related to the local 60-$\mu$m luminosity function by simple luminosity
evolution.  We assume a Gaussian luminosity function with no evolution over some specified
redshift range.  Our objects, when evolved to, $z=0$ are early-type stellar populations, 
quite unrelated to the objects which dominate the local 60-$\mu$m luminosity function.  
We consequently derive a far lower
normalization to our luminosity functions than do Blain et al.~(1999a).
The present observations do not permit us to distinguish 
between these models, but this should be a straightforward exercise when a 
significantly complete redshift distribution for the SCUBA sources is known. 

In addition, Eales et al.~(1998) argue that at least 10 percent
of the stars in the Universe
formed in SCUBA sources since they produce about
10 percent 
of the extragalactic background light.  This number is much closer to the
numbers presented in this paper.  Nevertheless, the assumption that the SCUBA sources
contribute the same fraction of the submillimeter background at all wavelengths, on which
the Eales et al.~calculation is based, appears to be inconsistent with the formulation that
we use, when the BIMA 2.8-mm counts, 
ISO 175-$\mu$m counts, and COBE 850-$\mu$m background
are all considered in conjuction.  Furthermore, it is unclear how we should relate
star-formation rates derived from optical or ultraviolet luminosities of normal field
galaxies to star-formation rates derived from submillimetre fluxes of the SCUBA sources.  

\begin{figure}
\begin{center}
\vskip -2mm
\epsfig{file=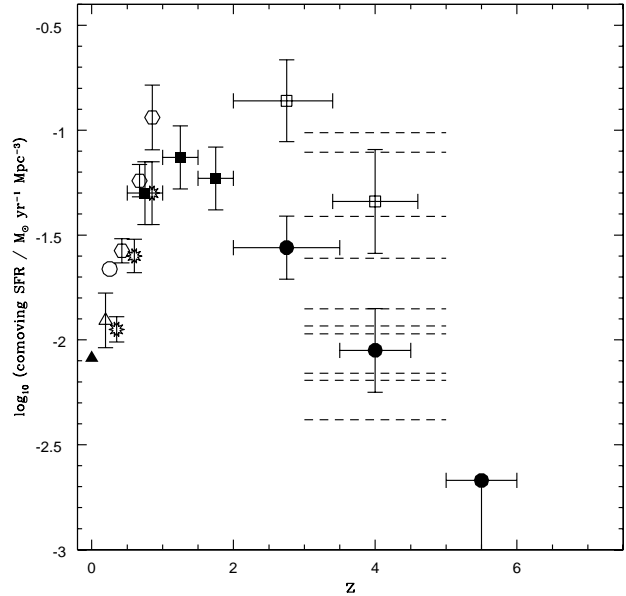, width=8.65cm}
\end{center}
\vskip -5mm
\caption{The comoving star-formation density of
the Universe contained in the SCUBA sources when their
luminosity-redshift distribution is as in Model B.
The eight dashed lines represent the SFR histories for models
(in ascending order) X, IX, IV, III, VI, V, VIII, VII, II and I.
The other points come from:
filled triangle -- Gallego et al.~(1996);
open triangle -- Treyer et al.~(1998);
open circle -- Tresse \& Maddox (1998);
stars -- Lilly et al.~(1996);
open hexagons -- Hammer \& Flores (1998);
filled squares - Connolly et al.~(1997);
filled circles - Madau et al.~(1996);
open squares - Pettini et al.~(1998; these include a global correction
for dust extinction).  The recent work of Glazebrook et al.~(1999) 
suggests a SFR almost coincident with the $z=0.85$ point of Hammer
\& Flores (1998).} 
\end{figure} 

\begin{figure}
\begin{center}
\vskip -2mm
\epsfig{file=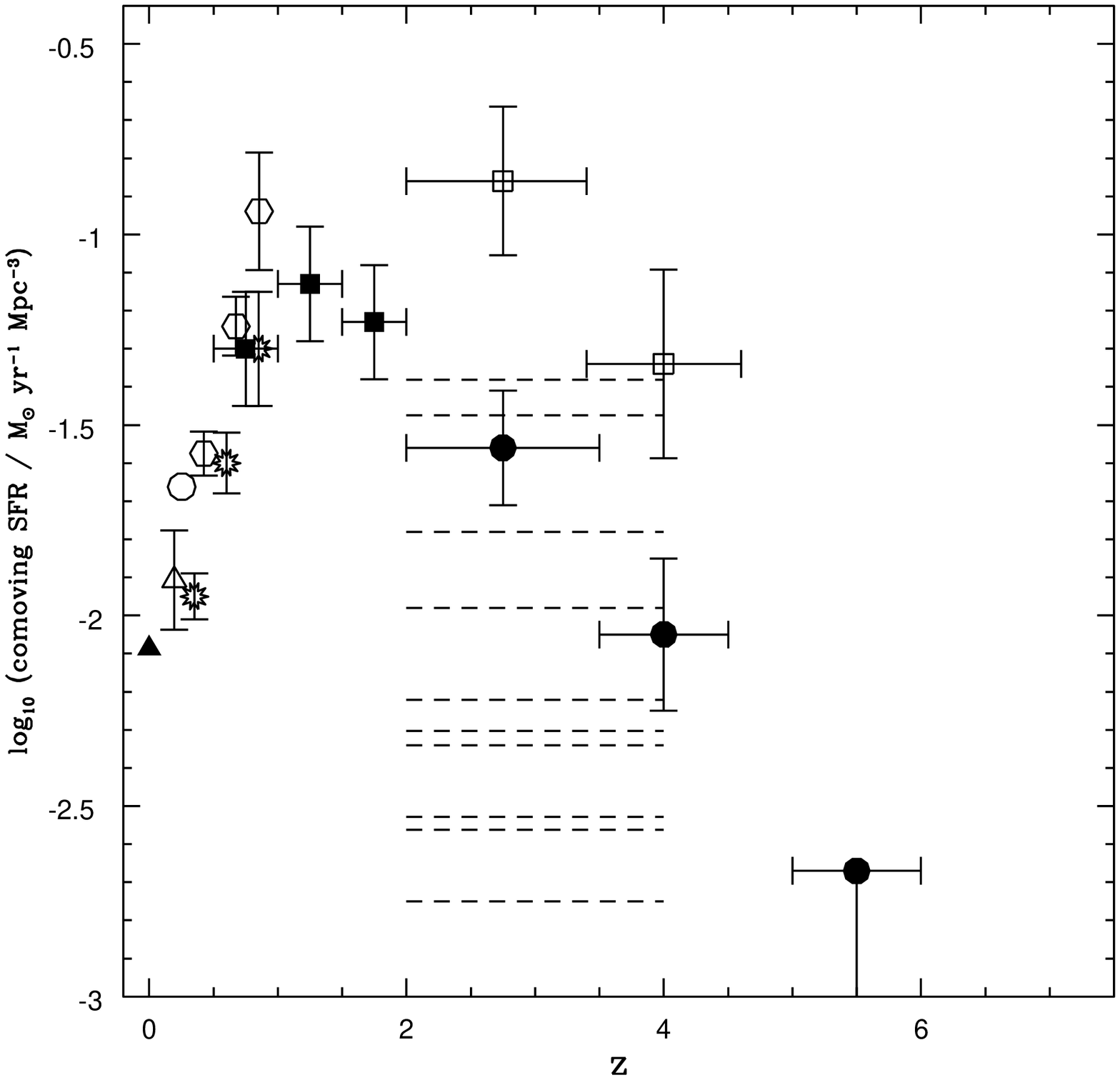, width=8.65cm}
\end{center}
\vskip-5mm
\caption{
As Figure 9, but for model F.  The symbols
have the same meanings and the lines are in the same order.
}
\end{figure} 

\begin{figure}
\begin{center}
\vskip-2mm
\epsfig{file=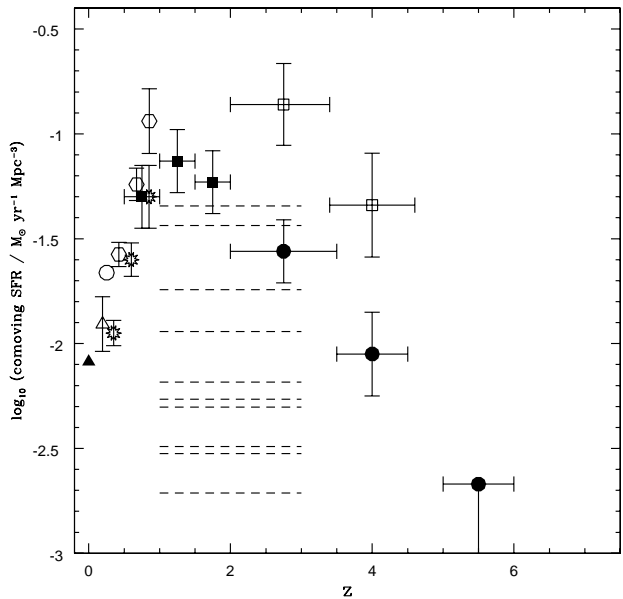, width=8.65cm}
\end{center}
\vskip-5mm
\caption{
As Figure 9, but for model G.  The symbols
have the same meanings and the lines are in the same order.
}
\end{figure} 

\subsubsection{The fate of the star-forming galaxies}

We have hinted that the SCUBA sources could evolve into elliptical galaxies.
Local ULIRGs have gas densities that are similar to the core stellar densities 
in elliptical galaxies (Kormendy \& Sanders 1992, Doyon et al.\ 1994).
If the SCUBA sources have similar morphologies to local ULIRGs, we 
might then expect them to evolve into elliptical galaxies.   

In Table 7 we present the density parameter in stars produced by the SCUBA 
sources, in our luminosity-function and star-formation models. For most models, 
particularly those with IMFs appropriate to elliptical galaxies (models IX and
X), these numbers are very small as compared with the stellar density 
contained in the local spheroid stellar population $\Omega_{\rm sph}$,
that is elliptical galaxies and bulges: $\Omega_{\rm sph} = 0.0036$ 
(Fukugita, Hogan \& Peebles 1998). 
In Models IX and X, between 1 and 4\,per cent of $\Omega_{\rm sph}$
formed in the luminous high-redshift SCUBA sources, and slightly more if a 
solar-neighbourhood IMF (Models III and IV) is assumed. These numbers are 
sufficiently low that a scenario in which all the SCUBA sources evolve into 
elliptical galaxies is consistent with galaxy formation models that predict
low-redshift elliptical formation (Kauffmann 1996; Kauffmann, Charlot \& White 
1996; Kauffmann \& Charlot 1998).  They are also low enough to be
consistent with the observed paucity of red galaxies in the Hubble Deep Field 
and the consequent interpretation that less than 30\,per cent of field ellipticals
formed at high redshift (Zepf 1997; Barger et al.\ 1998b).  

Various lines of argument suggest that the elliptical galaxies in clusters are 
very old. Not only did the stars form a long time ago (Ellis et al.\ 1997; 
Stanford et al.\ 1998; Kodama et al.\ 1998), but it appears that in the richest 
clusters the most luminous elliptical galaxies themselves were assembled by 
$z \sim 1$ (Trentham \& Mobasher 1998). It is further possible that most 3C radio 
galaxies are cluster ellipticals in the process of formation (Best, Longair \& 
Rottgering 1998). This leads to a natural question: are the SCUBA galaxies 
forming cluster ellipticals?  Low-redshift cluster ellipticals are
highly clustered, with a bias of about 4, and an even greater bias at high redshifts
(Fry 1996; Mo \& White 1996). In comparison, the reasonably uniform detection
rate of SCUBA galaxies in the fields studied by Smail et al.\ (1998) suggests 
that this is unlikely. 
These fields contained low-redshift clusters (which magnify the background
SCUBA sources through gravitational lensing), but these low-redshift
clusters are unrelated to the hypothesized ellipticals in formation that
we are discussing here. 
Instead, we propose that the SCUBA sources are a 
forming trace population of field ellipticals.  This is consistent with the 
constraints outlined in the previous paragraph.  This is not to say that if we 
happened to point SCUBA at a cluster in formation we would not detect a large 
number of galaxies.

\begin{table*} 
\caption{Comoving star-formation rates in units of 
M$_\odot$\,yr$^{-1}$\,Mpc$^{-3}$ in the three well fitting models of  
distant ULIRGs.
}
{$$\vbox{
\halign {\hfil #\hfil && \quad \hfil #\hfil \cr
\noalign{\hrule \medskip}
LF Model & SF I & SF II & SF III & SF IV & SF V & SF VI & SF VII & SF VIII & SF
IX & SF X &\cr
\noalign{\medskip \hrule \smallskip}
\cr
B & 0.097 & 0.078  & 0.011  & 0.0069 & 0.014  & 0.012  & 0.039 & 0.025 & 0.0064
& 0.0041 &\cr
F & 0.042 & 0.034  & 0.0046 & 0.0030 & 0.0060 & 0.0050 & 0.017 & 0.010 & 0.0027
& 0.0018 &\cr
G & 0.045 & 0.037 & 0.0050  & 0.0032 & 0.017 & 0.0054 & 0.018 & 0.011 & 0.0030 &
 0.0019 &\cr
\noalign{\medskip \hrule}
\noalign{\smallskip}\cr}}$$}
\end{table*}
 
\begin{table*} 
\caption{The present-day values of the density parameter in stars $\Omega_{*}$ 
produced by each well fitting 
luminosity function (LF) model, in each star-formation model.} 
{$$\vbox{
\halign {\hfil #\hfil && \quad \hfil #\hfil \cr
\noalign{\hrule \medskip}
LF Model & SF I & SF II & SF III & SF IV & SF V & SF VI & SF VII & SF VIII & SF
IX & SF X &\cr
\noalign{\medskip \hrule \smallskip}
\cr
B & 0.0010 & 0.00082 & 0.00012 & 7.3 $\times$ 10$^{-5}$  & 0.00015  & 0.00013 &
0.00041 & 0.00026 & 6.7 $\times$ 10$^{-5}$ & 4.3 $\times$ 10$^{-5}$ &\cr
F & 0.00080 & 0.00064 & 8.7 $\times$ 10$^{-5}$ & 5.7 $\times$ 10$^{-5}$ & 
0.00011
& 9.5 $\times$ 10$^{-5}$  & 0.00032  & 0.00019 & 5.1 $\times$ 10$^{-5}$ & 3.4
$\times$ 10$^{-5}$ &\cr
G & 0.0019 & 0.0016 & 0.00021 & 0.00014 & 0.00072  & 0.00023  & 0.00077 & 0.00047 
& 0.00013 & 8.0 $\times$ 10$^{-5}$ &\cr
\noalign{\medskip \hrule}
\noalign{\smallskip}\cr}}$$}
\end{table*}

\begin{table*} 
\caption{Cosmic enrichment from each well fitting model. The results are 
in units of the solar metallicity, assumed to be 0.0189 (Anders
\& Grevesse 1989). The redshift range is specified in the luminosity function 
model: see Table\,2. All the enrichment due to distant ULIRGs is 
completed by the lower limit to this redshift range.} 
{$$\vbox{
\halign {\hfil #\hfil && \quad \hfil #\hfil \cr
\noalign{\hrule \medskip}
LF Model & $z$ range & SF I & SF II & SF III & SF IV & SF V & SF VI &
SF VII & SF  VIII & SF IX & SF X &\cr
\noalign{\medskip \hrule \smallskip}
\cr
B & $3<z<5$ & 0.12 & 0.29 & 0.010 & 0.021 & 0.012 & 0.030  & 0.0023 & 0.0047 & 0
.010 & 0.021 &\cr
F & $2<z<4$ & 0.050 & 0.13 & 0.0044 & 0.0090 & 0.0051 & 0.012  & 0.0010 & 0.0020
 & 0.0044 & 0.0090 &\cr
G & $1<z<3$ & 0.055 & 0.14 & 0.0048 & 0.0098 & 0.0056 & 0.014  & 0.0011 & 0.0022
 & 0.0048 & 0.0097 &\cr
\noalign{\medskip \hrule}
\noalign{\smallskip}\cr}}$$}
\end{table*}

It has been recently recognized that lyman-break galaxies are highly
clustered at high redshift (Pettini et al.~1998), and so it is worth
investigating a possible link between these objects and the SCUBA sources.
Looking at the spectra of the two kinds of galaxies, we see that 
that lyman-break galaxies and the SCUBA sources are very likely two completely
separate populations of objects (for example compare the 0.5 $\mu$m
spectrum of the lyman-break galaxy 1512-cB58 
at $z=2.7$ taken by Pettini et al.~1998
using LRIS on Keck II with the 0.5 $\mu$m spectrum of the SCUBA source
SMM\,J02399$-$0136 at $z=2.8$
taken by Ivison et al.~1998 using the MOS spectrograph
on the CFHT).
However, while this observation is interesting, it does not rule out a
hypothetical link since
it is possible that one type of galaxy evolves into the other on
short timescales.
Measuring correlation functions would be the
best way of testing such a hypothetical link.  Unfortunately, the forty or
so SCUBA sources found so far 
do not have redshift measurements, so such an analysis is not
possible at present.  Such a calculation will, however, be possible once
a large sample of SCUBA sources with measured redshifts is available.

If the elliptical galaxies that went through a submillimetre-luminous
phase represent a sub-population, then their star-formation history 
must be similar to that of the rest of the remaining field
elliptical population. Otherwise a huge scatter in the stellar mass-to-light
versus light correlation (Oegerle \& Hoessel 1991) would be produced.

A few ellipticals (e.g. Stockton et al.\ 1995, Dunlop et al.\ 1996) must have formed
at high redshift and do not lie in very rich X-ray clusters. It is not obvious 
that these galaxies will evolve into cluster ellipticals.  The existence of such 
objects lends additional support to a scenario in which the SCUBA sources 
evolve into a trace sub-population of field elliptical galaxies. Furthermore, 
many high-redshift quasars are surrounded by heavily enriched gas, and so there 
must have been considerable star formation in their host galaxies at high 
redshift (Matteucci \& Padovani 1993). 
 
\subsubsection{Metal production}

In Table 8, we present the total amount of metals produced by the SCUBA 
sources for our models. The average metallicity of the Universe at $z=0$ is 
poorly known, but is probably about half solar -- see for example Section 5 of 
Fukugita et al.\ (1998) -- if the solar metallicity is 0.0189 (Anders \& Grevesse 
1989), and is much larger than the values listed in Table 8. Hence, it appears 
that a sufficiently small fraction of cosmic star-formation activity occurs in the 
SCUBA sources that they still may be a sub-population of elliptical galaxies 
undergoing a monolithic collapse at high redshift, without significantly 
overproducing the metal mass density of the Universe. 
On the other hand,
if {\it all} elliptical galaxies formed via monolithic collapse
at high redshift, then this would result in a metal mass density of the
Universe that is higher than that observed 
-- the ``cluster-field analogy'' of 
Madau 1997: see also Mushotzky \& Loewenstein (1997) and Renzini (1997).

For comparison, if the enrichment of the Ly$\alpha$ forest is regarded as a 
signature of cosmic enrichment, then the metallicity of the intergalactic 
medium at $z \sim 3$ is about 10$^{-2.5}$ to 10$^{-2}$ solar (Cowie et al.\ 1995;
Songalia \& Cowie 1996; Songaila 1997; Cowie \& Songaila 1998), and may be 
higher at lower redshift. 
This value is close to the metallicities in Table 8. 
But this similarity is perhaps not too meaningful because one would expect that
most of the metals produced in the SCUBA 
galaxies would remain bound within the galaxies.  This is
because ellipticals have deep 
gravitational potential wells and thus low ejection efficiencies (Nath \& Chiba 
1995); based on their gas consumption alone, the SCUBA sources must have 
masses in excess of 10$^{10}$\,M$_{\odot}$.  Hence, only a small fraction of the 
metals should leak out into the intergalactic medium, and so the values 
listed in Table 8 are best interpreted as
upper limits when comparing with the intergalactic
medium metallicity, if the SCUBA sources are all at redshifts greater than
three. 
We also note that most low-redshift elliptical galaxies have metal-poor 
interstellar media: this is not what we would expect if the metals produced in the
early history of these galaxies remained bound to the galaxy, as we argue above. 
One possible explanation of this apparent discrepancy is that 
the elliptical galaxies accumulate substantial X-ray halos (e.g.~Forman, Jones \& 
Tucker 1985), which then strip the metals out of the centres of galaxies.  Another 
possible explanation is that the SCUBA sources evolve into the few local 
ellipticals that {\it do} have metal-rich interstellar media, for example 
Centaurus A.    
 
\subsection{The SCUBA sources as AGN}

An alternative interpretation is that the SCUBA sources derive their bolometric 
luminosities from dust-enshrouded AGN and not from star formation at all.  
A number of lines of argument suggest that this may well be a plausible scenario.

First, the one SCUBA-selected source that has been studied in detail 
(SMM\,J02399$-$0136; Ivison et al.~1998) contains a powerful dust-enshrouded 
AGN. Note however that it is still possible for most of its bolometric luminosity 
to be produced in a starburst. The hyperluminous gravitationally lensed 
galaxy IRAS F10214+4724 at $z=2.29$ (Rowan-Robinson et al.~1991; Lacy 
et al.~1998 and references therein) also contains a dust-enshrouded AGN. 
The other two known ultraluminous high-redshift dusty galaxies H\,1413+117 
(Barvainis et al.\ 1995) and APM\,08279+5255 (Irwin et al.\ 1998; Lewis et al. 1998) 
also contain powerful AGN. If these objects are at all typical of 
high-redshift submillimetre-luminous sources, then it may well be plausible 
that a signficant fraction of their bolometric luminosities are powered by AGN as 
opposed to starbursts. 
 
A second line of argument is based on the observation that in local Universe, 
massive dark objects (MDOs) -- probably supermassive black holes -- are 
inferred to reside in the centers of most nearby galaxies (Magorrian et al.\ 1997), 
based on gas- and stellar-dynamical measurements (Kormendy \& Richstone 
1995). These MDOs are plausibly dead quasar engines that have run out of fuel.

Magorrian et al.\ (1997) show that the mass of the inferred MDO is well 
correlated with the luminosity of the spheroid stellar population in the host
galaxy. The total mass density in MDOs in the local Universe is about 
1.7 $\times$ 10$^6$\,M$_{\odot}$\,Mpc$^{-3}$ (Haehnelt et al.\ 1998), about an 
order of magnitude greater than the inferred mass density of black holes that is 
required to explain the blue light of quasars (Phinney 1998, Haehnelt et al.\
1998), unless the accretion efficiency in quasars is much less than 1\,per cent. 

One interesting idea suggested by Haehnelt et al.\ (1998) is that most 
accretion onto supermassive black holes in the centres of galaxies is obscured 
by dust.  If this is the case, then the presence of a population of dust-obscured 
AGN can explain the above discrepancy. The SCUBA sources would be 
natural candidates for this population.  We use our luminosity functions to 
calculate the comoving mass density of MDOs, $\rho_{\rm MDO}$ required to 
generate the observed bolometric luminosity of the SCUBA sources, assuming 
accretion at some efficiency $\epsilon$ -- this is the accretion luminosity 
divided by the Eddington luminosity:
\begin{equation} 
\rho_{\rm MDO} =  10^4 \alpha 
          \left( { {\epsilon} \over {0.1}}\right)^{-1}
          {\rm M}_{\odot} {\rm Mpc}^{-3}. 
\end{equation}
$\alpha = 3.8$, 1.6 and 1.8 in Models B, F and G respectively. 
These numbers are all below the local mass density of MDOs.  
Therefore it is plausible that all the SCUBA sources are AGN, but 
they may only help to explain the discrepancy between the local density
of MDOs and the inferred density of black holes required to explain the
blue light of quasars if their accretion luminosity is significantly
lower than 1\,per cent of the Eddington value. Recent work by Almaini, 
Lawrence \& Boyle (submitted), based on the X-ray luminosity function and 
background confirms that a large fraction of the SCUBA sources could be 
dust-enshrouded AGN. 

A third, but weaker, line of reasoning is based on the analogy with the 
ultraluminous $z=0.44$ cD galaxy IRAS\,P09104$-$4109 (Kleinmann et al.\ 1988), 
which with a bolometric luminosity of about $3 \times 10^{13}$\,L$_{\odot}$ is 
by far the most luminous object detected at mid-infrared wavelengths at  
$z<0.5$. It is at the centre of a very substantial cooling flow (Fabian \& Crawford 
1995). Two of the cD galaxies in the lensing clusters studied by Smail et al.\ 
(1998) were detected at 850\,$\mu$m (Edge et al. submitted), although they 
are not included in the counts presented by Smail et al.\ (1997;
1999). If the physical 
processes that make IRAS\,P09104$-$4109 into a hyperluminous galaxy are 
related to its being in the center of a cluster, then we might expect the same 
processes in the cDs detected by SCUBA.  The bolometric luminosity of 
IRAS P09104$-$4109 appears to come from a powerful dust-enshrouded AGN 
(Kleinmann et al.\ 1988), and an analogous mechanism could plausibly be 
powering the high-redshift SCUBA sources.

\section{Limitations of the present work and future prospects} 

The main limitations of the present work are that: 
\begin{enumerate}
\item we cannot determine whether the SCUBA sources are star-forming 
galaxies or AGN. Clearly the implications for cosmology would be very 
different for these two scenarios; 
\item we cannot determine uniquely the redshift-dependent luminosity function 
of the SCUBA sources, either to tell which of Models B, F and G or a 
different kind of model (Blain et al.\ 1999a) are most correct; 
\item we cannot determine the conversion factor to transform the far-infrared 
luminosity into a star-formation rate for the SCUBA sources to within one 
order of magnitude, even if all the far-infrared luminosity is assumed to be 
generated by star formation.
\end{enumerate} 

Given these limitations, we suggest the following future studies to address 
each of the problems listed above: 
\begin{enumerate}
\item Optical spectroscopic observations of SCUBA galaxies should be made to 
search for AGN, for example to determine the line ratios of hydrogen 
recombination to forbidden lines (Sanders 1998b; Veilleux et al.\ 1995; 
Evans 1996) and carry out spectral polarimetry (e.g.~Hines \& Wills 1993,
Januzzi et al.\ 1994). Note, however, that these measurements are extremely 
difficult to carry out for very distant galaxies, and cannot be used to determine 
unambiguously whether most of the bolometric luminosity comes from a 
starburst or an AGN, because of the huge amount of internal extinction
towards the galaxy centres at optical wavelengths. Genzel et al.\ (1998) used 
the ratio of high- to low-excitation mid-infrared emission lines in combination 
with the strength of a 7.7-$\mu$m PAH feature to search for AGN in local ULIRGs, 
but it will be quite some time before such measurements become available for the 
distant SCUBA sources. Crude interpretations can be made on the basis of 
source temperatures and SEDs (Sanders et al.\ 1998a), 
for example compare the SEDs of Arp 220 and Mrk 231.
Perhaps the best we can hope for in the near future is to match the
SED of each SCUBA source to that of a plausible
low-redshift counterpart, for which mid-infrared observations have
been used to determine a power source, and assume the high-redshift source 
is powered in the same way; 
\item an accurate determination of the redshift-dependent luminosity function 
will become available once redshifts for a significant number of SCUBA sources 
are obtained.  In addition, further surveys with SCUBA and other 
millimetre/submillimetre-wave detectors (Glenn et al.\ 1998) should 
allow the luminosity function to be determined accurately (Blain 1998).
For many of our models (Table 8), the total star-formation rate in high-redshift 
ULIRGs is greater than at low redshifts; a measurement of the redshift distribution 
would allow this statement to be quantified far more accurately.
\item selecting star-formation parameters to use in the Leitherer \& Heckman 
models is a difficult exercise.  This is a particular problem for the distant
SCUBA galaxies, for which only very limited information about
their global properties is available. Progress in this area would be possible if
the low-redshift remnants of the SCUBA galaxies were identified, and 
their stellar populations studied in detail in order to investigate their IMF.
Some clues are given in Section 3.1, but the conclusions there are far
from rigorous at this stage. 
\end{enumerate} 

More generally, placing the SCUBA galaxies in the context of the
galaxy-formation problem is clearly an important exercise. Measurements of the 
redshift distribution will be an crucial part of this.  For example, these 
measurements will let us see where the comoving bolometric luminosity density 
of the SCUBA sources peaks, relative to the peak in the comoving 
star-formation rate in other galaxies (Madau et al.\ 1996, Madau et al.\ 1998, 
Pettini et al.\ 1998) and to the peak in the comoving AGN luminosity density at 
ultraviolet (Boyle \& Terlevich 1998) or radio (Dunlop 1998) wavelengths. 

\section{Conclusions}

\begin{enumerate}
\item We identify various redshift-dependent luminosity functions that are 
consistent with the 850-$\mu$m SCUBA source counts and observations 
at other far-infrared and submillimetre wavelengths. Consistency can only be 
achieved if the 850-$\mu$m SCUBA source counts are towards the lower end 
of the range suggested by Smail et al.\ (1997).
In all of the consistent models, the SCUBA sources have high bolometric 
luminosities in excess of 
$10^{12}$\,L$_{\odot}$.
\item At present we cannot identify whether the power sources in the SCUBA
galaxies are predominantly starbursts or AGN. 
\item If the SCUBA sources are dusty ultraluminous star-forming galaxies, then 
they need not dominate the star-formation rate of the Universe at any redshift.  
In particular,
if they evolve into elliptical galaxies, our models suggest that they only 
contribute a few percent of the local spheroidal stellar population. 
This result is consistent with a scenario in which most elliptical
galaxies formed at low redshift, as is implied by the paucity of red
galaxies in wide-field $K$-band surveys. 
\item The alternative scenario in which the SCUBA sources are powerful 
dust-enshrouded AGN is also possible.  If the AGN are powered by accretion 
onto a black hole with a luminosity that is about one per cent of the Eddington 
value, then they may help to explain away the discrepancy of a factor of ten 
between the observed mass density of supermassive black holes in the local 
Universe and that required to produce the blue light of high-redshift quasars.
\item Tight constraints on the redshift-dependent luminosity function of
submillimetre-luminous galaxies at high redshift will follow from spectroscopic 
measurements of a large sample of SCUBA sources (Barger et al. in preparation). 
Obtaining accurate star-formation rates for the galaxies and identifying the 
nature of the power source will be more difficult. Matching the SEDs of 
the SCUBA sources to those of local ULIRGs, for which power sources have 
been determined using mid-infrared spectroscopy, offers one way to make 
progress in this regard. 
\end{enumerate} 

\section*{Acknowledgments} 

We thank our colleagues in Cambridge and Hawaii for many helpful 
discussions and the PPARC for financial support. We thank an 
anonymous referee for his/her prompt and helpful comments.


\begin{thebibliography}{}

\bibitem[\protect\citename{bl}%
]{ALB}
Almaini O., Lawrence A., Boyle B.\,J., 1999, MNRAS, submitted

\bibitem[\protect\citename{bl}%
]{An}
Anders E., Grevesse N., 1989, Geochimica et Cosmochimica, 53, 197 

\bibitem[\protect\citename{bl}%
]{Ar}
Armus L., Neugebauer G., Soifer B.\,T., Matthews K., 
1995, AJ, 110, 2610

\bibitem[\protect\citename{bl}%
]{Au}
Aussel H., Cesarsky C.\,J., Elbaz D., Starck J.\,-L., 
1998, A\&A, in press (astro-ph/9810044)  

\bibitem[\protect\citename{bl}%
]{Ba}
Barger A.\,J., Cowie L.\,L., Sanders D.\,B., Fulton E., Taniguchi Y., 
Sato Y., Kawara K., Okuda H., 1998a, Nat, 394, 248 

\bibitem[\protect\citename{bl}%
]{Ba2}
Barger A.\,J., Cowie L.\,L., Trentham N., Fulton E., Hu E.\,M., 
Songaila A., Hall D., 1998b, AJ, submitted


\bibitem[protectcitename{blah}%
]{Ba}
Barvainis R., Antonucci R., Hurt T., Coleman P., Reuter H.-P., 1995, ApJ,
451, L9

\bibitem[\protect\citename{bl}%
]{Be}
Benford D.\,J., Cox P., Omont A., Phillipps T.\,G., 1998, BAAS, 30, 783

\bibitem[\protect\citename{bl}%
]{Best}
Best P.\,N., Longair M.\,S., Rottgering H.\,J.\,A., 1998, MNRAS, 294, 549

\bibitem[\protect\citename{bl}%
]{Bi}
Binggeli B., Sandage A., Tamman G.\,A., 1988, ARA\&A, 26, 509

\bibitem[\protect\citename{bl}%
]{BL98}
Blain A.\,W., 1998, in Colombi S., Mellier Y. eds, 
in Wide-field surveys in cosmology. Proc. XIV IAP meeting, 
Editions Fronti\`eres, 
Gif-sur-Yvette, in press (astro-ph/9806369)

\bibitem[\protect\citename{bl}%
]{bl}
Blain A.\,W., Longair M.\,S., 1993, MNRAS, 264, 509 

\bibitem[\protect\citename{bl}%
]{bsik}
Blain A.\,W., Smail I., Ivison R.\,J., Kneib J.\,-P., 1999a, MNRAS, in press 
(astro-ph/9806062) 

\bibitem[\protect\citename{bl}%
]{B99}
Blain A.\,W., Kneib J.-P., Ivison R.\,J., Smail I., 1999b, ApJ, submitted

\bibitem[\protect\citename{bl}%
]{bt}
Boyle B.\,J., Terlevich R., 1998, MNRAS, 293, L49

\bibitem[\protect\citename{bl}%
]{c}
Connolly A.\,J., Szalay A.\,S., Dickinson M., Subbarao M.\,U.,
Brunner R.\,J., 1997, ApJ, 486, L11 

\bibitem[\protect\citename{bl}%
]{cs}
Cowie L.\,L., Songaila A., 1998, Nat, 394, 44

\bibitem[\protect\citename{bl}%
]{csk}
Cowie L.\,L., Songaila A., Kim T.-S., Hu E.\,M., 1995, AJ, 109, 1522

\bibitem[]{Downes}
Downes D., Neri R., Wiklind T., Wilner D.\,J., Shaver P., ApJL, submitted
(astro-ph/9810111)

\bibitem[\protect\citename{bl}%
]{do}
Doyon R., Wells M., Wright G.\,S., Joseph R.\,D., Nadeau D.,
James P.\,A., 1994, ApJ, 437, L23

\bibitem[\protect\citename{bl}%
]{du} 
Dunlop J., Peacock J., Spinrad H., Dey A., Jimenez R., Stern D.,
Windhorst R., 1996, Nat, 381, 581

\bibitem[\protect\citename{bl}%
]{dun}
Dunlop J.\,S., 1998, in Bremer M.\,N., Jackson N., Perez-Fournon I., eds, 
Observational Cosmology with the New Radio Surveys.~Kluwer, Dordrecht, p.~157 
(astro-ph/9704294)

\bibitem[\protect\citename{bl}%
]{dw}
Dwek E.\ et al., 1998, ApJ, in press (astro-ph/9806129)

\bibitem[\protect\citename{bl}%
]{EE1} 
Eales S.\,A., Edmunds M.\,G., 1996, MNRAS, 280, 1167

\bibitem[\protect\citename{bl}%
]{EE2} 
Eales S.\,A., Edmunds M.\,G., 1997, MNRAS, 286, 732

\bibitem[\protect\citename{bl}%
]{EL}
Eales S.\,A., Lilly S.\,J., Gear W.\,K., Bond J.\,R., Dunne L., Hammer F., 
Le F\`evre O., Crampton D., 1998, ApJL, submitted
(astro-ph/9808040)

\bibitem[\protect\citename{bl}%
]{Edge}
Edge A., Ivison R.\,J., Smail I., Blain A.\,W., Kneib J.-P., MNRAS, 
submitted

\bibitem[\protect\citename{bl}%
]{Ell}
Ellis R.\,S., Smail I., Dressler A., Couch W.\,J., Oemler A., Butcher H.,
Sharples R.\,M., 1997, ApJ, 483, 582

\bibitem[\protect\citename{bl}%
]{Ev} 
Evans A.\,S., 1996, PhD Thesis, University of Hawaii

\bibitem[\protect\citename{bl}%
]{Faber}
Faber S.\,M., 1973, ApJ, 179, 731

\bibitem[\protect\citename{bl}%
]{FC}
Fabian A.\,C., Crawford C.\,S., 1995, MNRAS, 274, L63

\bibitem[\protect\citename{bl}%
]{FS} 
Ferguson H.\,C., Sandage A., 1991, AJ, 101, 765

\bibitem[\protect\citename{bl}%
]{Fix}
Fixsen D.\,J., Dwek E., Mather J.\,C., Bennett C.\,L., Shafter R.\,A.,
1998, ApJ, in press (astro-ph/9803021) 

\bibitem[\protect\citename{bl}%
]{FJ} 
Forman W., Jones C., Tucker W., 1985, ApJ, 293, 102

\bibitem[\protect\citename{bl}%
]{FAD}
Franceschini A., Andreani P., Danese L., 1998, MNRAS, 296, 709

\bibitem[\protect\citename{bl}%
]{Fetal} 
Frayer D.\,T., Ivison R.\,J., Scoville N.\,Z., Yun M.\,S., Evans A.\,S., Smail I., 
Blain A.\,W., Kneib J.-P., 1998, ApJ, 506, L7 (astro-ph/9808109)

\bibitem[\protect\citename{bl}%
]{Fr}
Fry J.\,N., 1996, ApJ, 461, L65  

\bibitem[\protect\citename{bl}%
]{Fu} 
Fukugita M., Hogan C.\,J., Peebles P.\,J.\,E., 1998, preprint
(astro-ph/9712020)

\bibitem[\protect\citename{bl}%
]{Ga}
Gallego J., Zamorano J., Arag{\'{o}}n-Salamanca A., 
Rego M., 1996, ApJ, 459, L43

\bibitem[\protect\citename{bl}%
]{ge} 
Genzel R.\ et al., 1998, ApJ, 498, 579

\bibitem[\protect\citename{bl}%
]{glazebrook}
Glazebrook K., Blake C., Economou F., Lilly S., Colless M., 1999, MNRAS, 
submitted (astro-ph/9808276)

\bibitem[\protect\citename{bl}%
]{Glenn}
Glenn J. et al., 1998, in Phillips T.\,G. ed., Advanced Technology
MMW, Radio and Terahertz telescopes. Proc. SPIE vol. 3357, SPIE, Bellingham, 
in press

\bibitem[\protect\citename{bl}%
]{Go1} 
Goldader J.\,D., Joseph R.\,D., Doyon R., Sanders D.\,B., 1995, ApJ, 44, 97

\bibitem[\protect\citename{bl}%
]{Go2} 
Goldader J.\,D., Joseph R.\,D., Doyon R., Sanders D.\,B., 1997a, ApJ, 474, 104

\bibitem[\protect\citename{bl}%
]{Go3} 
Goldader J.\,D., Joseph R.\,D., Doyon R., Sanders D.\,B., 1997b, ApJS, 108, 449

\bibitem[\protect\citename{blah}%
]{Guiderdoni}
Guiderdoni B., Bouchet F.\,R., Puget J.-L., Lagache G., Hivon E., 1997,
Nat, 390, 257

\bibitem[\protect\citename{bl}%
]{HNR} 
Haehnelt M.\,G., Natarajan P., Rees M.\,J., 1998, MNRAS, in press 
(astro-ph/9712259) 

\bibitem[\protect\citename{bl}%
]{HF} 
Hammer F., Flores H., 1998, in 
Thuan T.\,X.\ et al., eds, Dwarf galaxies 
and cosmology: 
Proc.~XVIII Moriond meeting, Editions Fronti\`eres, Gif-sur-Yvette, in press 
(astro-ph/9806184)

\bibitem[\protect\citename{bl}%
]{Ha} 
Hauser M.\,G.\ et al., 1998, ApJ, in press (astro-ph/9806167) 

\bibitem[\protect\citename{bl}%
]{HW} 
Hines D.\,C., Wills B.\,J., 1993, ApJ, 415, 82

\bibitem[\protect\citename{bl}%
]{Ho1} 
Holland W.\,S. et al. 1999, MNRAS, in press (astro-ph/9809122)

\bibitem[\protect\citename{bl}%
]{Ho2} 
Holland W.\,S. et al., 1998, Nat, 392, 788 

\bibitem[\protect\citename{bl}%
]{Hu1}  
Hughes D.\,et al., 1998, Nat, 394, 241 

\bibitem[\protect\citename{bl}%
]{Hu2} 
Hughes D., 1996, in Bremer M.\,N., van der Werf P.\,P., 
Rottgering H.\,J.\,A., Carilli C.\,L., ed.,  
Cold Gas at High Redshift.~Kluwer, Dordrecht, p.~311

\bibitem[\protect\citename{bl}%
]{Hu1} 
Hunter D.\,A., Shaya E.\,J., Holtzman J.\,A., Light R.\,M.,
O'Neil E.\,J., Lynds R., 1995, ApJ, 448, 179

\bibitem[\protect\citename{bl}%
]{Hu2} 
Hunter D.\,A., Light R.\,M., Holtzman J.\,A., Lynds R., O'Neil E.\,J.,
Grillmair C.\,J., 1997, ApJ, 478, 124

\bibitem[\protect\citename{bl}%
]{Ietal}
Irwin M.\,J., Ibata R.\,A., Lewis G.\,F., Totten E.\,J., ApJ, 505, 529
(astro-ph/9806171)

\bibitem[\protect\citename{bl}%
]{I+7} 
Ivison R., Smail I., Le Borgne J.-F., Blain A.\,W., Kneib J.-P.,
Bezecourt J., Kerr T.\,H., Davies J.\,K., 1998, MNRAS, 298, 583 

\bibitem[\protect\citename{bl}%
]{J} 
Januzzi B.\,T., Elston R., Schmidt G.\,D., Smith P.\,S., Stockman H.\,S.,
1994, ApJ, 429, L49

\bibitem[\protect\citename{bl}%
]{K1} 
Kauffmann G., 1996, MNRAS, 281, 487

\bibitem[\protect\citename{bl}%
]{K2} 
Kauffmann G., Charlot S., 1998, MNRAS, 297, L23

\bibitem[\protect\citename{bl}%
]{K3} 
Kauffmann G., Charlot S., White S.\,D.\,M., 1996, MNRAS, 283, L117

\bibitem[\protect\citename{bl}%
]{Ka} 
Kawara K.\ et al., 1997, in Wilson A., ed., The Far-Infrared
and Submillimetre Universe.~ESA publications, 
Noordwijk, p.\,285  

\bibitem[\protect\citename{bl}%
]{Ka2}
Kawara K.\ et al., 1998, A\&A, 336, L9

\bibitem[\protect\citename{bl}%
]{Ki} 
Kim D.-C., Sanders D.\,B., Veilleux S., Mazzarella J.\,M., 
Soifer B.\,T., 1995, ApJS, 98, 171

\bibitem[\protect\citename{bl}%
]{Kl} 
Kleinmann S.\,G., Hamilton D., Keel W.\,C., Wynn-Williams C.\,G.,
Eales S.\,A., Becklin E.\,E., Kuntz K.\,D., 1988, ApJ, 328, 161 

\bibitem[\protect\citename{bl}%
]{Ko} 
Kodama T., Arimoto N., Barger A.\,J., Aragon-Salam{\`{a}}nca A., 1998,
A\&A, 334, 99

\bibitem[\protect\citename{bl}%
]{KR} 
Kormendy J., Richstone D., 1995, ARAA, 33, 581 

\bibitem[\protect\citename{bl}%
]{KS} 
Kormendy J., Sanders D.\,B., 1992, ApJ, 390, L53

\bibitem[\protect\citename{bl}%
]{L} 
Lacy M., Rawlings S., Serjeant S., 1998, MNRAS, in press (astro-ph/9806079) 

\bibitem[\protect\citename{bl}%
]{La} 
Lagache G.\ et al., 1998, in Colombi S., Mellier Y., eds, 
Wide-field surveys in cosmology, Proc. XIV IAP 
Conference.~ Editions Fronti\`eres, Gif-Sur-Yvette, in press

\bibitem[\protect\citename{bl}%
]{La} 
Larkin J.\,E., Armus L., Knop K., Matthews K., Soifer B.\,T., 
1995, ApJ, 452, 599

\bibitem[\protect\citename{bl}%
]{Le} 
Leitherer C., 1998, in Gilmore G.,ed., The 38th
Herstmonceux Conference: The Stellar Initial Mass Function.~ASP,
San Frncisco, in press   

\bibitem[\protect\citename{bl}%
]{LH} 
Leitherer C., Heckman T.\,M., 1995, ApJS, 96, 9

\bibitem[\protect\citename{Lewis et al.}]{Lewis}
Lewis G.\,F., Chapman S.\,C., Ibata R.\,A., Irwin M.\,J., Totten E.\,J., 1998, 
ApJ, 505, L1

\bibitem[\protect\citename{bl}%
]{Li} 
Lilly S.\,J., Le F{\`{e}}vre O., Hammer F., Crampton D., 1996, ApJ, 460, L1

\bibitem[\protect\citename{bl}%
]{LE} 
Lilly S.\,J., Eales S.\,A., Gear W.\,K., Bond J.\,R., Dunne L., Hammer F., 
Le F\`evre O., Crampton D., 1998, in Benvenuti J. et al.~eds., 
NGST: Science Drivers and
Technical Challenges.~ESA publications, Noordwijk, in press (astro-ph/9807261)

\bibitem[\protect\citename{bl}%
]{L} 
Lutz D.\ et al., 1996, A\&A, 315, L137

\bibitem[\protect\citename{bl}%
]{MF} 
Madau P., Ferguson H.\,C., Dickinson M.\,E., Giavalisco M., Steidel C.\,C.,
Fruchter A., 1996, MNRAS, 293, 1388  

\bibitem[\protect\citename{bl}%
]{M} 
Madau P., Della Valle M., Panagia N., 1998, MNRAS, 297, L17 

\bibitem[\protect\citename{bl}%
]{Ma} 
Magorrian J.\ et al., 1998, AJ, 115, 2285 

\bibitem[\protect\citename{bl}%
]{MJ} 
Massey P., Johnson K.\,E., DeGioia-Eastwood K., 1995a, ApJ, 454, 151

\bibitem[\protect\citename{bl}%
]{ML} 
Massey P., Lang C.\,C., DeGioia-Eastwood K., Garmany C.\,D., 1995b, 
ApJ, 438, 188

\bibitem[\protect\citename{bl}%
]{Ma} 
Matteucci F., Padovani P., 1994, ApJ, 419, 485

\bibitem[\protect\citename{bl}%
]{Me} 
Meurer G.\,R., Heckman T.\,M., Leitherer C., Kinney A., Robert C.,
Garnett D.\,R., 1995, AJ, 110, 2665

\bibitem[\protect\citename{bl}%
]{MW} 
Mo H.\,J., White S.\,D.\,M., 1996, MNRAS, 282, 1096 

\bibitem[\protect\citename{bl}%
]{MT} 
Mouri H., Taniguchi Y., 1992, ApJ, 386, 68

\bibitem[\protect\citename{bl}%
]{ML}
Mushotzky R.\,F., Loewenstein M., 1997, ApJ, 481, L63

\bibitem[\protect\citename{bl}%
]{NC} 
Nath B.\,B., Chiba M., 1995, ApJ, 454 604

\bibitem[\protect\citename{bl}%
]{OH} 
Oegerle W.\,R., Hoessel J.\,G., 1991, ApJ, 375, 15 

\bibitem[\protect\citename{bl}%
]{P}
Pettini M., Steidel C.\,C., Adelberger K.\,L., Kellogg M.,
Dickinson M., Giavalisco M., 1998, in
Shull J.\,M., Woodward C.\,E., Thronson H.\,A., eds., 
Cosmic Origins: evolution of 
galaxies, stars, planets, and life.~Astr.~Soc.~Pac., 
San Francisco, in press (astro-ph/9708117)

\bibitem[\protect\citename{bl}%
]{Ph} 
Phinney E.\,S., 1998, in Barnes J., Sanders D., eds., IAU Symposium 186:
Galaxy Interactions at Low and High Redshift.~Kluwer, Dordrecht, in press    

\bibitem[\protect\citename{bl}%
]{PJ} 
Prestwich A.\,H., Joseph R.\,D., Wright G.\,S., 1994, ApJ, 422, 73

\bibitem[\protect\citename{bl}%
]{Pu} 
Puget J.-L., Abergel A., Bernard J.-P., Boulanger F., Burton W.~B.,
D{\'{e}}sert F.-X., Hartmann D., 1996, A\&A, 308, L5

\bibitem[\protect\citename{bl}%
]{Re} 
Renzini A., 1997, ApJ, 488, 35 

\bibitem[\protect\citename{bl}%
]{RR1}  
Rowan-Robinson M.\ et al., 1991, Nat, 351, 719
 
\bibitem[\protect\citename{bl}%
]{RR2}  
Rowan-Robinson M.\ et al., 1997, MNRAS, 289, 490

\bibitem[\protect\citename{bl}%
]{S1} 
Sanders D.\,B., Soifer B.\,T., Elias J.\,H., Madore B.\,F., Matthews K.,
Neugebauer G., Scoville N.\,Z., 1988a, ApJ, 325, 74

\bibitem[\protect\citename{bl}%
]{S2} 
Sanders D.\,B., Soifer B.\,T., Elias J.\,H., Neugebauer G., Matthews K.,
1998b, ApJ, 328, L35

\bibitem[\protect\citename{bl}%
]{SM} 
Sanders D.\,B, Mirabel I.\,F., 1996, ARA\&A, 34, 749

\bibitem[\protect\citename{bl}%
]{Sa} 
Saunders W., Rowan-Robinson M., Lawrence A., Efstathiou G., Kaiser N.,
Ellis R.\,S., Frenk C.\,S., 1990, MNRAS, 242, 318

\bibitem[\protect\citename{bl}%
]{Sch} 
Schlegel D.\,J., Finkbeiner D.\,P., Davis M., 1998, ApJ, 500, 525 

\bibitem[\protect\citename{bl}%
]{Sc}
Scoville N.\,Z.\ et al., 1998, ApJ, 492, L107

\bibitem[\protect\citename{bl}%
]{SIB} 
Smail I., Ivison R.\,J., Blain A.\,W., 1997, ApJ, 490, L5 

\bibitem[\protect\citename{bl}%
]{SIBK} 
Smail I., Ivison R.\,J., Blain A.\,W., Kneib J.-P., 1998, ApJ, 507, L21
(astro-ph/9806061) 

\bibitem[\protect\citename{bl}%
]{SIBKII} 
Smail I., Ivison R., Blain A., Kneib J.-P., 1999, in Holt S.\,S., Smith E.\,P. 
eds, After the dark ages: when galaxies were young. AIP, Woodbury NY, in press 
(astro-ph/9810281)

\bibitem[\protect\citename{bl}%
]{So1} 
Songaila A., 1997, ApJ, 490, L1

\bibitem[\protect\citename{bl}%
]{SC} 
Songaila A., Cowie L.\,L., 1996, AJ, 112, 335 

\bibitem[\protect\citename{bl}%
]{St} 
Stanford S.\,A., Eisenhardt P.\,R., Dickinson M., 1998, ApJ, 492, 461

\bibitem[\protect\citename{bl}%
]{Sto}
Stockton A., Kellogg M., Ridgway S.\,E., 1995, ApJ, 443, L69 

\bibitem[\protect\citename{bl}%
]{T1} 
Trentham N., Kormendy J., Sanders D.\,B., 1998, AJ, submitted

\bibitem[\protect\citename{bl}%
]{TM}  
Trentham N., Mobasher B., 1998, MNRAS, in press (astro-ph/9805282) 

\bibitem[\protect\citename{bl}%
]{TrMa} 
Tresse L., Maddox S.\,J., 1998, ApJ, 495, 691

\bibitem[\protect\citename{bl}%
]{T} 
Treyer M.\,A., Ellis R.\,S., Milliard B., Donas J., Bridges T.\,J., 
1998, MNRAS, in 
press (astro-ph/9806056) 

\bibitem[\protect\citename{bl}%
]{V} 
Vader P., 1986, ApJ, 305, 1669 

\bibitem[\protect\citename{bl}%
]{Ve} 
Veilleux S., Kim D.-C., Sanders D.\,B., Mazzarella J.\,M., Soifer B.\,T., 
1995, ApJS, 98, 171

\bibitem[\protect\citename{bl}%
]{WW} 
Wilner D.\,J., Wright M.\,C.\,H., 1997, ApJ, 488, L67

\bibitem[\protect\citename{bl}%
]{Z} 
Zepf S., 1997, Nat, 390, 377

\bibitem[\protect\citename{bl}%
]{ZS} 
Zepf S., Silk J., 1996, ApJ, 466, 114

\end{thebibliography}
\end{document}